\documentclass[aps,prc,reprint,superscriptaddress,nofootinbib,amssymb,amsmath,amsfonts]{revtex4-2}

\usepackage[dvipsnames]{xcolor}
\usepackage[utf8]{inputenc}
\usepackage{physics} 
\usepackage{graphicx} 
\usepackage{booktabs} 
\usepackage{multirow} 
\usepackage{verbatim} 
\usepackage{longtable}
\usepackage{array}
\newcolumntype{C}{>{$}c<{$}}
\newcolumntype{L}{>{$}l<{$}}
\renewcommand{\arraystretch}{1.25}

\usepackage{hyperref}
\usepackage[capitalise]{cleveref} 

\begin{document}
\title{Minimal complete sets for two pseudoscalar meson photoproduction}

\author{Philipp Kroenert}
\author{Yannick Wunderlich}
\author{Farah Afzal}
\author{Annika Thiel}
\affiliation{Helmholtz Institut für Strahlen- und Kernphysik, Universität Bonn, Germany}

\begin{abstract}
    For photoproduction reactions with final states consisting of two pseudoscalar mesons and a spin-1/2 baryon, 8 complex amplitudes need to be determined uniquely.
    A modified version of Moravcsik's theorem is employed for these reactions, resulting in slightly over-complete sets of polarization observables that are able to determine the amplitudes uniquely.
    Further steps were taken to reduce the found sets to minimal complete sets.
    As a final result, multiple minimal complete sets without any remaining ambiguities are presented for the first time.
    These sets consist of $2N=16$ observables, containing only one triple polarization observable.
\end{abstract}

\maketitle

\section{Introduction}
The interrelation between experiment and theory is what drives science.
In the field of hadron spectroscopy these are the measurement of cross sections or polarization observables and its counterpart Quantum Chromodynamics.
The latter describes the transition of the initial to the final state via a transition matrix $\mathcal{T}$.
This matrix comprises the employed model predictions to describe a certain process.
Via so-called formation experiments (i.e. $\gamma p \rightarrow \pi\pi p$) it is possible to study the emergence of resonant states (such as $\Delta(1232)$, $N(1440)1/2+$, $N(1520)3/2-$, etc. \cite{Sokhoyan}).

These states can be analyzed via partial wave analysis (i.e. BnGa \cite{Sokhoyan}, MAID \cite{fix2005double}), determining the matrix elements of $\mathcal{T}$ and comparing it to the model prediction.
However, as polarization observables depend on bilinear products of the complex amplitudes \cite{FasanoTabakin,Knoechlein,RobertsOed}, mathematical ambiguities arise \cite{ChiangTabakin}.
Nevertheless, it is still possible to determine unique solutions by employing a complete experiment analysis \cite{BDS}.

Such a complete experiment analysis was performed analytically by Chiang and Tabakin in 1997 \cite{ChiangTabakin} for single pseudoscalar meson photoproduction.
A detailed proof comprising all the relevant cases was published recently by Nakayama \cite{Nakayama}.
It should be noted that these complete experiments are an idealization for data with no uncertainty \cite{Ireland2010}.
Although the process of single pseudoscalar meson photoproduction can be fully described by only four complex amplitudes \cite{cgln}, the calculations are non trivial and cumbersome \cite{Nakayama} and furthermore, quite involved ambiguity-structures can arise.

Within this paper, the determination of complete sets of observables is studied for the reaction of two pseudoscalar meson photoproduction.
The process can be described by $N=8$ complex amplitudes and thus allows for 64 measurable polarization observables \cite{RobertsOed}, which are four times as many observables as in the case of single pseudoscalar meson photoproduction \cite{ChiangTabakin}.
This results in an exponential increase of complexity, for what reason the algebraic techniques presented in \cite{Nakayama} are no longer appropriate, although possible (see \cref{sec::algebraic_method}).
New methods should be employed, in order to allow for an easier access to the problem of complete sets for reactions with $N>4$.

There is an already existing work on this subject by Arenhövel and Fix \cite{PhysRevC.89.034003} from 2014.
On the one hand, they used the inverse function theorem to derive complete sets of 16 observables.
The downside of this method is that the resulting sets might locally be free of ambiguities, but not globally.
On the other hand, they used a graph theoretical approach, where a complex amplitude is represented as a node and the bilinear product as a connection between certain nodes.
This method yields complete sets with 25 observables.
It was then shown how to further reduce such a set to 15 observables.
Although, they found sets without triple polarization observables, there still remain quadratic ambiguities.

To overcome these difficulties arising from remaining discrete mathematical ambiguities, Moravcsik's theorem \cite{Moravcsik} is employed within this paper.
This theorem allows for the extraction of complete sets of observables for an arbitrary number of amplitudes.
Furthermore, due to its graph-theoretical foundation, the whole algorithm can be automated \cite{otherPaperOfUs}.

The paper is structured in the following way:
Starting point is a short recap of Moravcsik's theorem and its modification in \cref{sec:M_theorem}.
\Cref{sec:2PiFormalism} introduces the 64 polarization observables for two pseudoscalar meson photoproduction.
\Cref{sec:database} elaborates on the difficulties of their experimental determination and gives an extensive overview of already performed measurements in two pseudoscalar meson photoproduction.
Within \cref{sec:approach} the actual application of Moravcsik's theorem is described and illustrated with an example.
The entire analysis results in 5964 unique, but slightly over-complete sets of observables.
Their characteristics are discussed in \cref{sec:results}.
In \cref{sec:reduction} it is described how to transform the slightly over-complete sets into minimal ones (i.e. into sets containing $2N = 16$ observables).
Based on these sets as well as the already performed measurements (see \cref{table:trulyMinimalSetsNoTriplePola,tab:collectionMeasurements1,tab:collectionMeasurements2}) the most promising minimal complete set is presented in \cref{sec:experimentalImplications}.
The results are summarized in \cref{sec:summary}.

\clearpage
\section{Moravcsik's theorem\label{sec:M_theorem}}
\begin{figure}
    \includegraphics[scale=0.6]{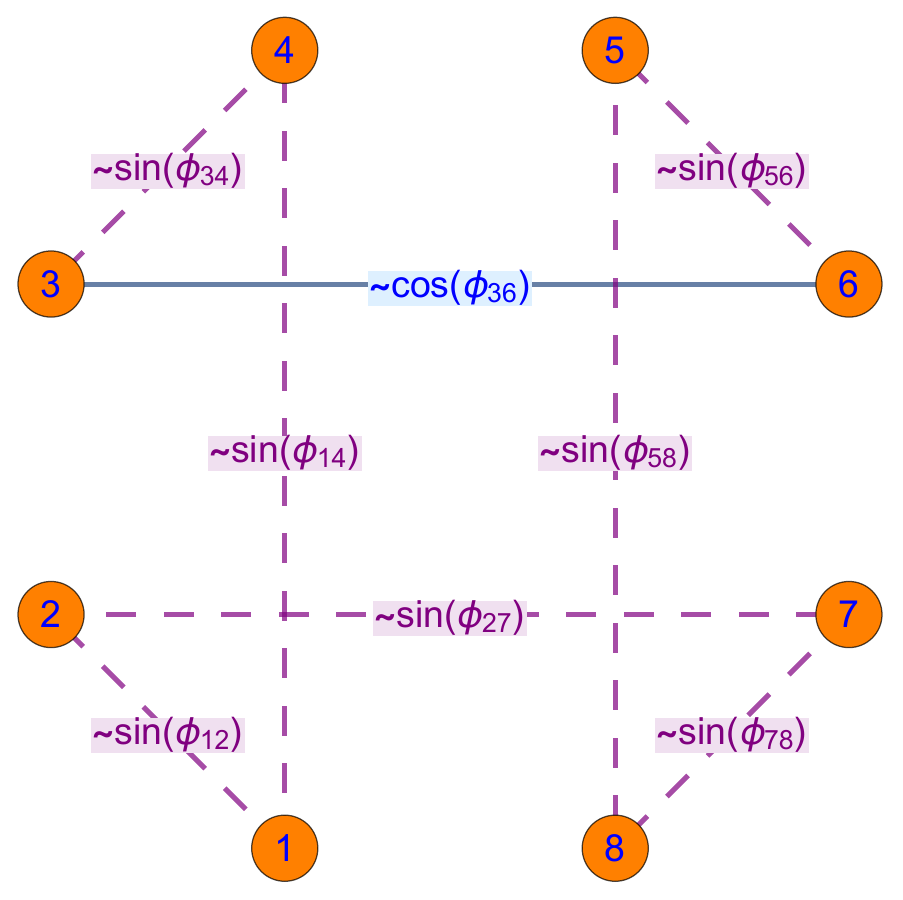}
    \caption{Illustration of a cycle graph with eight nodes (enumerated points) and edges (solid and dashed lines). Each node represents one complex amplitude, whereas each edge connecting two nodes represents either the real (solid) or imaginary (dashed) part of the bilinear product of those nodes. The respective correlation to the relative phase $\phi_{ij}$ is indicated. \label{fig:graph_example}}
\end{figure}

The main points of Moravcsik's paper \cite{Moravcsik} shall be recapped in a concise form.
The basic assumption of the theorem is that the moduli of the $N$ complex amplitudes $t_i$ are known, together with the real and imaginary parts of the bilinear products $t_i t^*_j$.
Furthermore, each complex amplitude $t_i$ is treated as a node of a graph whereas an edge is the real/imaginary part of the bilinear amplitude product $t_i t^*_j$ connecting both nodes.
An illustration is shown in \cref{fig:graph_example}.
Such a graph is said to correspond to a complete set of observables if it fulfills the following two requirements:
\begin{enumerate}
    \item it is a connected graph
    \item it has an odd number of edges which corresponds to an imaginary part of a particular bilinear product, i.e. $\sim\Im(t_i t_{j}^{*})$
\end{enumerate}
The first condition is related to the “consistency relation" of the relative phases:
\begin{align}
    \phi_{12} + \phi_{23} + \ldots + \phi_{N1} = 0, \label{eqn:consistency relation}
\end{align}
which implies a summation of relative phases between all neighboring amplitudes $t_i$ \cite{Nakayama}.
\Cref{eqn:consistency relation} has to hold in every case, whether the considered set is fully complete, or not.
The second condition is responsible for resolving the discrete ambiguities since it holds
\begin{align}
    \Im(t_i t^*_j) = |t_i||t_j| \cdot \sin(\phi_{ij}),
\end{align}
and that sine itself produces an ambiguity due to its periodicity:
\begin{align}
    \phi_{ij} \rightarrow (\phi_{ij}, \pi - \phi_{ij}).
\end{align}
It turns out that any odd number of such “sine-type" ambiguities resolves the discrete ambiguities, due to the summand $\pi$.
The generalization to any odd number is the actual modification to Moravcsik's theorem.
A proof of the original version of the theorem can be found in \cite{Moravcsik} and a quite detailed proof of the modified version of the theorem is given in reference \cite{otherPaperOfUs}.

The following analysis focuses on cycle graphs, i.e. connected graphs where each node has degree two.
As explained in the paper \cite{Moravcsik}: “from the point of view of eliminating discrete ambiguities" these graph types are “the most economical" ones.
Thus, only the minimal number of $N$ bilinear products is needed in order to eliminate all discrete ambiguities, one for each edge.

\section{Polarization observables\label{sec:2PiFormalism}}
The derivation of the $64$ polarization observables of two pseudoscalar meson photoproduction was first published by Roberts and Oed \cite{RobertsOed}.
The observables were defined in a "helicity and hybrid helicity-transversity basis" \cite{RobertsOed}.
For the latter, the photon spin is still quantized along its direction of motion.
For the sake of comparability, the hybrid basis shall be adopted in this paper.
However, in order to work out the connection between the real (imaginary) part of the bilinear products and the relative phases $\phi_{ij}$, it is advantageous to rename the amplitudes:
\begin{align}
b_1^+ \rightarrow t_1 && b_2^+ \rightarrow t_2 && b_3^+ \rightarrow t_3 && b_4^+ \rightarrow t_4\\
b_1^- \rightarrow t_5 && b_2^- \rightarrow t_6 && b_3^- \rightarrow t_7 && b_4^- \rightarrow t_8
\end{align}
Within reference \cite{RobertsOed}, the observables are ordered according to the polarization of the photon beam which is required to measure the respective observable.
This ordering scheme is advantageous from an experimental point of view, unfortunately it is inappropriate when studying ambiguities.
Therefore, the observables are regrouped according to their mathematical structure, which yields eight groups.
While the first group consists of observables solely described by the squared moduli of the amplitudes $t_i$, any other group comprises equal amounts of observables containing only $\cos$- or $\sin$-terms.
The resulting expressions for the observables are listed in \cref{observables1,observables2}.

For the purpose of an easier calculation, 64 $\Gamma$-matrices are introduced, which can be solely described by the identity matrix, as well as the three Pauli matrices (listed in \cref{definition_gamma_matrices}).
Allowing to calculate the respective observable by the bilinear form $\ev{\Gamma}{t}$ with $\ket{t}:=(t_1,t_2,t_3,t_4,t_5,t_6,t_7,t_8)$, similar as in \cite{ChiangTabakin}.
As expected, the $\Gamma$-matrices for each group share the same matrix structure.
Naturally, they form an orthogonal basis, are hermitian and unitary.
Indeed, the matrices fulfill the same properties as presented in \cite{ChiangTabakin} (with an adapted prefactor in the orthogonality relation).

\begin{table*}[h]
\caption{Definitions of the 64 polarization observables for two pseudoscalar meson photoproduction in hybrid helicity-transversity form. Here, $\phi_{ij}$ denotes the relative phase between the complex amplitudes $t_i$ and $t_j$. The notation used in the original paper of Roberts and Oed \cite{RobertsOed} is also shown. The observables are classified into eight groups according to their underlying mathematical structure. The vector $\ket{t}$ has the form $(t_1,t_2,t_3,t_4,t_5,t_6,t_7,t_8)$ and the shape of the $\Gamma$-matrices is outlined in \cref{definition_gamma_matrices}.\label{observables1}}
\begin{ruledtabular}
\begin{tabular}{CLCC}
    \text{Observable} & \multicolumn{1}{C}{\text{Definition in terms of polar coordinates / 2}} & \text{Bilinear form} & \text{Roberts/Oed}\\
    \midrule
    \mathcal{O}^{\text{I}}_{1} & 2\cdot(|t_1| ^2+|t_2| ^2+|t_3| ^2+|t_4| ^2-|t_5| ^2-|t_6| ^2-|t_7| ^2-|t_8| ^2) & \ev{\Gamma^{\text{I}}_{1}}{t} & \text{I}^{\odot } \\
    \mathcal{O}^{\text{I}}_{2} & 2\cdot(|t_1| ^2+|t_2| ^2-|t_3| ^2-|t_4| ^2+|t_5| ^2+|t_6| ^2-|t_7| ^2-|t_8| ^2) & \ev{\Gamma^{\text{I}}_{2}}{t} & \text{P}_{\text{y}} \\
    \mathcal{O}^{\text{I}}_{3} & 2\cdot(|t_1| ^2-|t_2| ^2+|t_3| ^2-|t_4| ^2+|t_5| ^2-|t_6| ^2+|t_7| ^2-|t_8| ^2) & \ev{\Gamma^{\text{I}}_{3}}{t} & \text{P}_{\text{y'}} \\
    \mathcal{O}^{\text{I}}_{4} & 2\cdot(|t_1| ^2-|t_2| ^2-|t_3| ^2+|t_4| ^2-|t_5| ^2+|t_6| ^2+|t_7| ^2-|t_8| ^2) & \ev{\Gamma^{\text{I}}_{4}}{t} & \mathcal{O}_{\text{yy'}}^{\odot} \\
    \mathcal{O}^{\text{I}}_{5} & 2\cdot(|t_1| ^2-|t_2| ^2-|t_3| ^2+|t_4| ^2+|t_5| ^2-|t_6| ^2-|t_7| ^2+|t_8| ^2) & \ev{\Gamma^{\text{I}}_{5}}{t} & \mathcal{O}_{\text{yy'}} \\
    \mathcal{O}^{\text{I}}_{6} & 2\cdot(|t_1| ^2-|t_2| ^2+|t_3| ^2-|t_4| ^2-|t_5| ^2+|t_6| ^2-|t_7| ^2+|t_8| ^2) & \ev{\Gamma^{\text{I}}_{6}}{t} & \text{P}_{\text{y'}}^{\odot } \\
    \mathcal{O}^{\text{I}}_{7} & 2\cdot(|t_1| ^2+|t_2| ^2-|t_3| ^2-|t_4| ^2-|t_5| ^2-|t_6| ^2+|t_7| ^2+|t_8| ^2) & \ev{\Gamma^{\text{I}}_{7}}{t} & \text{P}_{\text{y}}^{\odot } \\
    \mathcal{O}^{\text{I}}_{8} & 2\cdot(|t_1| ^2+|t_2| ^2+|t_3| ^2+|t_4| ^2+|t_5| ^2+|t_6| ^2+|t_7| ^2+|t_8| ^2) & \ev{\Gamma^{\text{I}}_{8}}{t} & \text{I}_0 \\
    \midrule
    \mathcal{O}^{\text{II}}_{\text{c1}} & \left|t_1\right|  \left|t_3\right|  \cos (\text{$\phi_{13}$})+\left|t_2\right|  \left|t_4\right|  \cos (\text{$\phi_{24}$})+\left|t_5\right|  \left|t_7\right|  \cos (\text{$\phi_{57}$})+\left|t_6\right|  \left|t_8\right|  \cos (\text{$\phi_{68}$}) & \ev{\Gamma ^{\text{II}}_{\text{c1}}}{t} & -\text{P}_{\text{z}} \\
    \mathcal{O}^{\text{II}}_{\text{c2}} & \left|t_1\right|  \left|t_3\right|  \cos (\text{$\phi_{13}$})+\left|t_2\right|  \left|t_4\right|  \cos (\text{$\phi_{24}$})-\left|t_5\right|  \left|t_7\right|  \cos (\text{$\phi_{57}$})-\left|t_6\right|  \left|t_8\right|  \cos (\text{$\phi_{68}$}) & \ev{\Gamma ^{\text{II}}_{\text{c2}}}{t} & -\text{P}_{\text{z}}^{\odot } \\
    \mathcal{O}^{\text{II}}_{\text{c3}} & \left|t_1\right|  \left|t_3\right|  \cos (\text{$\phi_{13}$})-\left|t_2\right|  \left|t_4\right|  \cos (\text{$\phi_{24}$})+\left|t_5\right|  \left|t_7\right|  \cos (\text{$\phi_{57}$})-\left|t_6\right|  \left|t_8\right|  \cos (\text{$\phi_{68}$}) & \ev{\Gamma ^{\text{II}}_{\text{c3}}}{t} & -\mathcal{O}_{\text{zy'}} \\
    \mathcal{O}^{\text{II}}_{\text{c4}} & \left|t_1\right|  \left|t_3\right|  \cos (\text{$\phi_{13}$})-\left|t_2\right|  \left|t_4\right|  \cos (\text{$\phi_{24}$})-\left|t_5\right|  \left|t_7\right|  \cos (\text{$\phi_{57}$})+\left|t_6\right|  \left|t_8\right|  \cos (\text{$\phi_{68}$}) & \ev{\Gamma ^{\text{II}}_{\text{c4}}}{t} & -\mathcal{O}_{\text{zy'}}^{\odot } \\
    \mathcal{O}^{\text{II}}_{\text{s1}} & \left|t_1\right|  \left|t_3\right|  \sin (\text{$\phi_{13}$})+\left|t_2\right|  \left|t_4\right|  \sin (\text{$\phi_{24}$})+\left|t_5\right|  \left|t_7\right|  \sin (\text{$\phi_{57}$})+\left|t_6\right|  \left|t_8\right|  \sin (\text{$\phi_{68}$}) & \ev{\Gamma ^{\text{II}}_{\text{s1}}}{t} & -\text{P}_{\text{x}} \\
    \mathcal{O}^{\text{II}}_{\text{s2}} & \left|t_1\right|  \left|t_3\right|  \sin (\text{$\phi_{13}$})+\left|t_2\right|  \left|t_4\right|  \sin (\text{$\phi_{24}$})-\left|t_5\right|  \left|t_7\right|  \sin (\text{$\phi_{57}$})-\left|t_6\right|  \left|t_8\right|  \sin (\text{$\phi_{68}$}) & \ev{\Gamma ^{\text{II}}_{\text{s2}}}{t} & -\text{P}_{\text{x}}^{\odot } \\
    \mathcal{O}^{\text{II}}_{\text{s3}} & \left|t_1\right|  \left|t_3\right|  \sin (\text{$\phi_{13}$})-\left|t_2\right|  \left|t_4\right|  \sin (\text{$\phi_{24}$})+\left|t_5\right|  \left|t_7\right|  \sin (\text{$\phi_{57}$})-\left|t_6\right|  \left|t_8\right|  \sin (\text{$\phi_{68}$}) & \ev{\Gamma ^{\text{II}}_{\text{s3}}}{t} & -\mathcal{O}_{\text{xy'}} \\
    \mathcal{O}^{\text{II}}_{\text{s4}} & \left|t_1\right|  \left|t_3\right|  \sin (\text{$\phi_{13}$})-\left|t_2\right|  \left|t_4\right|  \sin (\text{$\phi_{24}$})-\left|t_5\right|  \left|t_7\right|  \sin (\text{$\phi_{57}$})+\left|t_6\right|  \left|t_8\right|  \sin (\text{$\phi_{68}$}) & \ev{\Gamma ^{\text{II}}_{\text{s4}}}{t} & -\mathcal{O}_{\text{xy'}}^{\odot } \\
    \midrule
    \mathcal{O}^{\text{III}}_{\text{c1}} & \left|t_1\right|  \left|t_2\right|  \cos (\text{$\phi_{12}$})+\left|t_3\right|  \left|t_4\right|  \cos (\text{$\phi_{34}$})+\left|t_5\right|  \left|t_6\right|  \cos (\text{$\phi_{56}$})+\left|t_7\right|  \left|t_8\right|  \cos (\text{$\phi_{78}$}) & \ev{\Gamma^{\text{III}}_{\text{c1}}}{t} & -\text{P}_{\text{z'}} \\
    \mathcal{O}^{\text{III}}_{\text{c2}} & \left|t_1\right|  \left|t_2\right|  \cos (\text{$\phi_{12}$})+\left|t_3\right|  \left|t_4\right|  \cos (\text{$\phi_{34}$})-\left|t_5\right|  \left|t_6\right|  \cos (\text{$\phi_{56}$})-\left|t_7\right|  \left|t_8\right|  \cos (\text{$\phi_{78}$}) & \ev{\Gamma^{\text{III}}_{\text{c2}}}{t} & -\text{P}_{\text{z'}}^{\odot } \\
    \mathcal{O}^{\text{III}}_{\text{c3}} & \left|t_1\right|  \left|t_2\right|  \cos (\text{$\phi_{12}$})-\left|t_3\right|  \left|t_4\right|  \cos (\text{$\phi_{34}$})+\left|t_5\right|  \left|t_6\right|  \cos (\text{$\phi_{56}$})-\left|t_7\right|  \left|t_8\right|  \cos (\text{$\phi_{78}$}) & \ev{\Gamma^{\text{III}}_{\text{c3}}}{t} & -\mathcal{O}_{\text{yz'}} \\
    \mathcal{O}^{\text{III}}_{\text{c4}} & \left|t_1\right|  \left|t_2\right|  \cos (\text{$\phi_{12}$})-\left|t_3\right|  \left|t_4\right|  \cos (\text{$\phi_{34}$})-\left|t_5\right|  \left|t_6\right|  \cos (\text{$\phi_{56}$})+\left|t_7\right|  \left|t_8\right|  \cos (\text{$\phi_{78}$}) & \ev{\Gamma^{\text{III}}_{\text{c4}}}{t} & -\mathcal{O}_{\text{yz'}}^{\odot } \\
    \mathcal{O}^{\text{III}}_{\text{s1}} & \left|t_1\right|  \left|t_2\right|  \sin (\text{$\phi_{12}$})+\left|t_3\right|  \left|t_4\right|  \sin (\text{$\phi_{34}$})+\left|t_5\right|  \left|t_6\right|  \sin (\text{$\phi_{56}$})+\left|t_7\right|  \left|t_8\right|  \sin (\text{$\phi_{78}$}) & \ev{\Gamma^{\text{III}}_{\text{s1}}}{t} & \text{P}_{\text{x'}} \\
    \mathcal{O}^{\text{III}}_{\text{s2}} & \left|t_1\right|  \left|t_2\right|  \sin (\text{$\phi_{12}$})+\left|t_3\right|  \left|t_4\right|  \sin (\text{$\phi_{34}$})-\left|t_5\right|  \left|t_6\right|  \sin (\text{$\phi_{56}$})-\left|t_7\right|  \left|t_8\right|  \sin (\text{$\phi_{78}$}) & \ev{\Gamma^{\text{III}}_{\text{s2}}}{t} & \text{P}_{\text{x'}}^{\odot } \\
    \mathcal{O}^{\text{III}}_{\text{s3}} & \left|t_1\right|  \left|t_2\right|  \sin (\text{$\phi_{12}$})-\left|t_3\right|  \left|t_4\right|  \sin (\text{$\phi_{34}$})+\left|t_5\right|  \left|t_6\right|  \sin (\text{$\phi_{56}$})-\left|t_7\right|  \left|t_8\right|  \sin (\text{$\phi_{78}$}) & \ev{\Gamma^{\text{III}}_{\text{s3}}}{t} & \mathcal{O}_{\text{yx'}} \\
    \mathcal{O}^{\text{III}}_{\text{s4}} & \left|t_1\right|  \left|t_2\right|  \sin (\text{$\phi_{12}$})-\left|t_3\right|  \left|t_4\right|  \sin (\text{$\phi_{34}$})-\left|t_5\right|  \left|t_6\right|  \sin (\text{$\phi_{56}$})+\left|t_7\right|  \left|t_8\right|  \sin (\text{$\phi_{78}$}) & \ev{\Gamma^{\text{III}}_{\text{s4}}}{t} & \mathcal{O}_{\text{yx'}}^{\odot } \\
    \midrule
    \mathcal{O}^{\text{IV}}_{\text{c1}} & \left|t_1\right|  \left|t_4\right|  \cos (\text{$\phi_{14}$})+\left|t_2\right|  \left|t_3\right|  \cos (\text{$\phi_{23}$})+\left|t_5\right|  \left|t_8\right|  \cos (\text{$\phi_{58}$})+\left|t_6\right|  \left|t_7\right|  \cos (\text{$\phi_{67}$}) & \ev{\Gamma^{\text{IV}}_{\text{c1}}}{t} & \mathcal{O}_{\text{zz'}} \\
    \mathcal{O}^{\text{IV}}_{\text{c2}} & \left|t_1\right|  \left|t_4\right|  \cos (\text{$\phi_{14}$})+\left|t_2\right|  \left|t_3\right|  \cos (\text{$\phi_{23}$})-\left|t_5\right|  \left|t_8\right|  \cos (\text{$\phi_{58}$})-\left|t_6\right|  \left|t_7\right|  \cos (\text{$\phi_{67}$}) & \ev{\Gamma^{\text{IV}}_{\text{c2}}}{t} & \mathcal{O}_{\text{zz'}}^{\odot } \\
    \mathcal{O}^{\text{IV}}_{\text{c3}} & \left|t_1\right|  \left|t_4\right|  \cos (\text{$\phi_{14}$})-\left|t_2\right|  \left|t_3\right|  \cos (\text{$\phi_{23}$})+\left|t_5\right|  \left|t_8\right|  \cos (\text{$\phi_{58}$})-\left|t_6\right|  \left|t_7\right|  \cos (\text{$\phi_{67}$}) & \ev{\Gamma^{\text{IV}}_{\text{c3}}}{t} & \mathcal{O}_{\text{xx'}} \\
    \mathcal{O}^{\text{IV}}_{\text{c4}} & \left|t_1\right|  \left|t_4\right|  \cos (\text{$\phi_{14}$})-\left|t_2\right|  \left|t_3\right|  \cos (\text{$\phi_{23}$})-\left|t_5\right|  \left|t_8\right|  \cos (\text{$\phi_{58}$})+\left|t_6\right|  \left|t_7\right|  \cos (\text{$\phi_{67}$}) & \ev{\Gamma^{\text{IV}}_{\text{c4}}}{t} & \mathcal{O}_{\text{xx'}}^{\odot } \\
    \mathcal{O}^{\text{IV}}_{\text{s1}} & \left|t_1\right|  \left|t_4\right|  \sin (\text{$\phi_{14}$})+\left|t_2\right|  \left|t_3\right|  \sin (\text{$\phi_{23}$})+\left|t_5\right|  \left|t_8\right|  \sin (\text{$\phi_{58}$})+\left|t_6\right|  \left|t_7\right|  \sin (\text{$\phi_{67}$}) & \ev{\Gamma^{\text{IV}}_{\text{s1}}}{t} & \mathcal{O}_{\text{xz'}} \\
    \mathcal{O}^{\text{IV}}_{\text{s2}} & \left|t_1\right|  \left|t_4\right|  \sin (\text{$\phi_{14}$})+\left|t_2\right|  \left|t_3\right|  \sin (\text{$\phi_{23}$})-\left|t_5\right|  \left|t_8\right|  \sin (\text{$\phi_{58}$})-\left|t_6\right|  \left|t_7\right|  \sin (\text{$\phi_{67}$}) & \ev{\Gamma^{\text{IV}}_{\text{s2}}}{t} & \mathcal{O}_{\text{xz'}}^{\odot } \\
    \mathcal{O}^{\text{IV}}_{\text{s3}} & \left|t_1\right|  \left|t_4\right|  \sin (\text{$\phi_{14}$})-\left|t_2\right|  \left|t_3\right|  \sin (\text{$\phi_{23}$})+\left|t_5\right|  \left|t_8\right|  \sin (\text{$\phi_{58}$})-\left|t_6\right|  \left|t_7\right|  \sin (\text{$\phi_{67}$}) & \ev{\Gamma^{\text{IV}}_{\text{s3}}}{t} & -\mathcal{O}_{\text{zx'}} \\
    \mathcal{O}^{\text{IV}}_{\text{s4}} & \left|t_1\right|  \left|t_4\right|  \sin (\text{$\phi_{14}$})-\left|t_2\right|  \left|t_3\right|  \sin (\text{$\phi_{23}$})-\left|t_5\right|  \left|t_8\right|  \sin (\text{$\phi_{58}$})+\left|t_6\right|  \left|t_7\right|  \sin (\text{$\phi_{67}$}) & \ev{\Gamma^{\text{IV}}_{\text{s4}}}{t} & -\mathcal{O}_{\text{zx'}}^{\odot }
\end{tabular}
\end{ruledtabular}
\end{table*}

\begin{table*}[h]
\caption{(Continuation of \cref{observables1}) Definitions of the 64 polarization observables for two pseudoscalar meson photoproduction in hybrid helicity-transversity form. Here, $\phi_{ij}$ denotes the relative phase between the complex amplitudes $t_i$ and $t_j$. The notation used in the original paper of Roberts and Oed \cite{RobertsOed} is also shown. The observables are classified into eight groups according to their underlying mathematical structure. The vector $\ket{t}$ has the form $(t_1,t_2,t_3,t_4,t_5,t_6,t_7,t_8)$ and the shape of the $\Gamma$-matrices is outlined in \cref{definition_gamma_matrices}.\label{observables2}}
\begin{ruledtabular}
\begin{tabular}{CLCC}
    \text{Observable} & \multicolumn{1}{C}{\text{Definition in terms of polar coordinates / 2}} & \text{Bilinear form} & \text{Roberts/Oed}\\
    \midrule
    \mathcal{O}^{\text{V}}_{\text{c1}} & \left|t_1\right|  \left|t_5\right|  \cos (\text{$\phi_{15}$})+\left|t_2\right|  \left|t_6\right|  \cos (\text{$\phi_{26}$})+\left|t_3\right|  \left|t_7\right|  \cos (\text{$\phi_{37}$})+\left|t_4\right|  \left|t_8\right|  \cos (\text{$\phi_{48}$}) & \ev{\Gamma ^{\text{V}}_{\text{c1}}}{t} & -\text{I}^{\text{c}} \\
    \mathcal{O}^{\text{V}}_{\text{c2}} & \left|t_1\right|  \left|t_5\right|  \cos (\text{$\phi_{15}$})+\left|t_2\right|  \left|t_6\right|  \cos (\text{$\phi_{26}$})-\left|t_3\right|  \left|t_7\right|  \cos (\text{$\phi_{37}$})-\left|t_4\right|  \left|t_8\right|  \cos (\text{$\phi_{48}$}) & \ev{\Gamma^{\text{V}}_{\text{c2}}}{t} & -\text{P}_{\text{y}}^{\text{c}} \\
    \mathcal{O}^{\text{V}}_{\text{c3}} & \left|t_1\right|  \left|t_5\right|  \cos (\text{$\phi_{15}$})-\left|t_2\right|  \left|t_6\right|  \cos (\text{$\phi_{26}$})+\left|t_3\right|  \left|t_7\right|  \cos (\text{$\phi_{37}$})-\left|t_4\right|  \left|t_8\right|  \cos (\text{$\phi_{48}$}) & \ev{\Gamma^{\text{V}}_{\text{c3}}}{t} & -\text{P}_{\text{y'}}^{\text{c}} \\
    \mathcal{O}^{\text{V}}_{\text{c4}} & \left|t_1\right|  \left|t_5\right|  \cos (\text{$\phi_{15}$})-\left|t_2\right|  \left|t_6\right|  \cos (\text{$\phi_{26}$})-\left|t_3\right|  \left|t_7\right|  \cos (\text{$\phi_{37}$})+\left|t_4\right|  \left|t_8\right|  \cos (\text{$\phi_{48}$}) & \ev{\Gamma^{\text{V}}_{\text{c4}}}{t} & -\mathcal{O}_{\text{yy'}}^{\text{c}} \\
    \mathcal{O}^{\text{V}}_{\text{s1}} & \left|t_1\right|  \left|t_5\right|  \sin (\text{$\phi_{15}$})+\left|t_2\right|  \left|t_6\right|  \sin (\text{$\phi_{26}$})+\left|t_3\right|  \left|t_7\right|  \sin (\text{$\phi_{37}$})+\left|t_4\right|  \left|t_8\right|  \sin (\text{$\phi_{48}$}) & \ev{\Gamma^{\text{V}}_{\text{s1}}}{t} & -\text{I}^{\text{s}} \\
    \mathcal{O}^{\text{V}}_{\text{s2}} & \left|t_1\right|  \left|t_5\right|  \sin (\text{$\phi_{15}$})+\left|t_2\right|  \left|t_6\right|  \sin (\text{$\phi_{26}$})-\left|t_3\right|  \left|t_7\right|  \sin (\text{$\phi_{37}$})-\left|t_4\right|  \left|t_8\right|  \sin (\text{$\phi_{48}$}) & \ev{\Gamma^{\text{V}}_{\text{s2}}}{t} & -\text{P}_{\text{y}}^{\text{s}} \\
    \mathcal{O}^{\text{V}}_{\text{s3}} & \left|t_1\right|  \left|t_5\right|  \sin (\text{$\phi_{15}$})-\left|t_2\right|  \left|t_6\right|  \sin (\text{$\phi_{26}$})+\left|t_3\right|  \left|t_7\right|  \sin (\text{$\phi_{37}$})-\left|t_4\right|  \left|t_8\right|  \sin (\text{$\phi_{48}$}) & \ev{\Gamma^{\text{V}}_{\text{s3}}}{t} & -\text{P}_{\text{y'}}^{\text{s}} \\
    \mathcal{O}^{\text{V}}_{\text{s4}} & \left|t_1\right|  \left|t_5\right|  \sin (\text{$\phi_{15}$})-\left|t_2\right|  \left|t_6\right|  \sin (\text{$\phi_{26}$})-\left|t_3\right|  \left|t_7\right|  \sin (\text{$\phi_{37}$})+\left|t_4\right|  \left|t_8\right|  \sin (\text{$\phi_{48}$}) & \ev{\Gamma^{\text{V}}_{\text{s4}}}{t} & -\mathcal{O}_{\text{yy'}}^{\text{s}} \\
    \midrule
    \mathcal{O}^{\text{VI}}_{\text{c1}} & \left|t_1\right|  \left|t_7\right|  \cos (\text{$\phi_{17}$})+\left|t_2\right|  \left|t_8\right|  \cos (\text{$\phi_{28}$})+\left|t_3\right|  \left|t_5\right|  \cos (\text{$\phi_{35}$})+\left|t_4\right|  \left|t_6\right|  \cos (\text{$\phi_{46}$}) & \ev{\Gamma^{\text{VI}}_{\text{c1}}}{t} & \text{P}_{\text{z}}^{\text{c}} \\
    \mathcal{O}^{\text{VI}}_{\text{c2}} & \left|t_1\right|  \left|t_7\right|  \cos (\text{$\phi_{17}$})+\left|t_2\right|  \left|t_8\right|  \cos (\text{$\phi_{28}$})-\left|t_3\right|  \left|t_5\right|  \cos (\text{$\phi_{35}$})-\left|t_4\right|  \left|t_6\right|  \cos (\text{$\phi_{46}$}) & \ev{\Gamma^{\text{VI}}_{\text{c2}}}{t} & -\text{P}_{\text{x}}^{\text{s}} \\
    \mathcal{O}^{\text{VI}}_{\text{c3}} & \left|t_1\right|  \left|t_7\right|  \cos (\text{$\phi_{17}$})-\left|t_2\right|  \left|t_8\right|  \cos (\text{$\phi_{28}$})+\left|t_3\right|  \left|t_5\right|  \cos (\text{$\phi_{35}$})-\left|t_4\right|  \left|t_6\right|  \cos (\text{$\phi_{46}$}) & \ev{\Gamma^{\text{VI}}_{\text{c3}}}{t} & \mathcal{O}_{\text{zy'}}^{\text{c}} \\
    \mathcal{O}^{\text{VI}}_{\text{c4}} & \left|t_1\right|  \left|t_7\right|  \cos (\text{$\phi_{17}$})-\left|t_2\right|  \left|t_8\right|  \cos (\text{$\phi_{28}$})-\left|t_3\right|  \left|t_5\right|  \cos (\text{$\phi_{35}$})+\left|t_4\right|  \left|t_6\right|  \cos (\text{$\phi_{46}$}) & \ev{\Gamma^{\text{VI}}_{\text{c4}}}{t} & -\mathcal{O}_{\text{xy'}}^{\text{s}} \\
    \mathcal{O}^{\text{VI}}_{\text{s1}} & \left|t_1\right|  \left|t_7\right|  \sin (\text{$\phi_{17}$})+\left|t_2\right|  \left|t_8\right|  \sin (\text{$\phi_{28}$})+\left|t_3\right|  \left|t_5\right|  \sin (\text{$\phi_{35}$})+\left|t_4\right|  \left|t_6\right|  \sin (\text{$\phi_{46}$}) & \ev{\Gamma^{\text{VI}}_{\text{s1}}}{t} & \text{P}_{\text{z}}^{\text{s}} \\
    \mathcal{O}^{\text{VI}}_{\text{s2}} & \left|t_1\right|  \left|t_7\right|  \sin (\text{$\phi_{17}$})+\left|t_2\right|  \left|t_8\right|  \sin (\text{$\phi_{28}$})-\left|t_3\right|  \left|t_5\right|  \sin (\text{$\phi_{35}$})-\left|t_4\right|  \left|t_6\right|  \sin (\text{$\phi_{46}$}) & \ev{\Gamma^{\text{VI}}_{\text{s2}}}{t} & \text{P}_{\text{x}}^{\text{c}} \\
    \mathcal{O}^{\text{VI}}_{\text{s3}} & \left|t_1\right|  \left|t_7\right|  \sin (\text{$\phi_{17}$})-\left|t_2\right|  \left|t_8\right|  \sin (\text{$\phi_{28}$})+\left|t_3\right|  \left|t_5\right|  \sin (\text{$\phi_{35}$})-\left|t_4\right|  \left|t_6\right|  \sin (\text{$\phi_{46}$}) & \ev{\Gamma^{\text{VI}}_{\text{s3}}}{t} & \mathcal{O}_{\text{zy'}}^{\text{s}} \\
    \mathcal{O}^{\text{VI}}_{\text{s4}} & \left|t_1\right|  \left|t_7\right|  \sin (\text{$\phi_{17}$})-\left|t_2\right|  \left|t_8\right|  \sin (\text{$\phi_{28}$})-\left|t_3\right|  \left|t_5\right|  \sin (\text{$\phi_{35}$})+\left|t_4\right|  \left|t_6\right|  \sin (\text{$\phi_{46}$}) & \ev{\Gamma^{\text{VI}}_{\text{s4}}}{t} & \mathcal{O}_{\text{xy'}}^{\text{c}} \\
    \midrule
    \mathcal{O}^{\text{VII}}_{\text{c1}} & \left|t_1\right|  \left|t_6\right|  \cos (\text{$\phi_{16}$})+\left|t_2\right|  \left|t_5\right|  \cos (\text{$\phi_{25}$})+\left|t_3\right|  \left|t_8\right|  \cos (\text{$\phi_{38}$})+\left|t_4\right|  \left|t_7\right|  \cos (\text{$\phi_{47}$}) & \ev{\Gamma^{\text{VII}}_{\text{c1}}}{t} & \text{P}_{\text{z'}}^{\text{c}} \\
    \mathcal{O}^{\text{VII}}_{\text{c2}} & \left|t_1\right|  \left|t_6\right|  \cos (\text{$\phi_{16}$})+\left|t_2\right|  \left|t_5\right|  \cos (\text{$\phi_{25}$})-\left|t_3\right|  \left|t_8\right|  \cos (\text{$\phi_{38}$})-\left|t_4\right|  \left|t_7\right|  \cos (\text{$\phi_{47}$}) & \ev{\Gamma^{\text{VII}}_{\text{c2}}}{t} & \mathcal{O}_{\text{yz'}}^{\text{c}} \\
    \mathcal{O}^{\text{VII}}_{\text{c3}} & \left|t_1\right|  \left|t_6\right|  \cos (\text{$\phi_{16}$})-\left|t_2\right|  \left|t_5\right|  \cos (\text{$\phi_{25}$})+\left|t_3\right|  \left|t_8\right|  \cos (\text{$\phi_{38}$})-\left|t_4\right|  \left|t_7\right|  \cos (\text{$\phi_{47}$}) & \ev{\Gamma^{\text{VII}}_{\text{c3}}}{t} & \text{P}_{\text{x'}}^{\text{s}} \\
    \mathcal{O}^{\text{VII}}_{\text{c4}} & \left|t_1\right|  \left|t_6\right|  \cos (\text{$\phi_{16}$})-\left|t_2\right|  \left|t_5\right|  \cos (\text{$\phi_{25}$})-\left|t_3\right|  \left|t_8\right|  \cos (\text{$\phi_{38}$})+\left|t_4\right|  \left|t_7\right|  \cos (\text{$\phi_{47}$}) & \ev{\Gamma^{\text{VII}}_{\text{c4}}}{t} & \mathcal{O}_{\text{yx'}}^{\text{s}} \\
    \mathcal{O}^{\text{VII}}_{\text{s1}} & \left|t_1\right|  \left|t_6\right|  \sin (\text{$\phi_{16}$})+\left|t_2\right|  \left|t_5\right|  \sin (\text{$\phi_{25}$})+\left|t_3\right|  \left|t_8\right|  \sin (\text{$\phi_{38}$})+\left|t_4\right|  \left|t_7\right|  \sin (\text{$\phi_{47}$}) & \ev{\Gamma^{\text{VII}}_{\text{s1}}}{t} & \text{P}_{\text{z'}}^{\text{s}} \\
    \mathcal{O}^{\text{VII}}_{\text{s2}} & \left|t_1\right|  \left|t_6\right|  \sin (\text{$\phi_{16}$})+\left|t_2\right|  \left|t_5\right|  \sin (\text{$\phi_{25}$})-\left|t_3\right|  \left|t_8\right|  \sin (\text{$\phi_{38}$})-\left|t_4\right|  \left|t_7\right|  \sin (\text{$\phi_{47}$}) & \ev{\Gamma^{\text{VII}}_{\text{s2}}}{t} & \mathcal{O}_{\text{yz'}}^{\text{s}} \\
    \mathcal{O}^{\text{VII}}_{\text{s3}} & \left|t_1\right|  \left|t_6\right|  \sin (\text{$\phi_{16}$})-\left|t_2\right|  \left|t_5\right|  \sin (\text{$\phi_{25}$})+\left|t_3\right|  \left|t_8\right|  \sin (\text{$\phi_{38}$})-\left|t_4\right|  \left|t_7\right|  \sin (\text{$\phi_{47}$}) & \ev{\Gamma^{\text{VII}}_{\text{s3}}}{t} & -\text{P}_{\text{x'}}^{\text{c}} \\
    \mathcal{O}^{\text{VII}}_{\text{s4}} & \left|t_1\right|  \left|t_6\right|  \sin (\text{$\phi_{16}$})-\left|t_2\right|  \left|t_5\right|  \sin (\text{$\phi_{25}$})-\left|t_3\right|  \left|t_8\right|  \sin (\text{$\phi_{38}$})+\left|t_4\right|  \left|t_7\right|  \sin (\text{$\phi_{47}$}) & \ev{\Gamma^{\text{VII}}_{\text{s4}}}{t} & -\mathcal{O}_{\text{yx'}}^{\text{c}} \\
    \midrule
    \mathcal{O}^{\text{VIII}}_{\text{c1}} & \left|t_1\right|  \left|t_8\right|  \cos (\text{$\phi_{18}$})+\left|t_2\right|  \left|t_7\right|  \cos (\text{$\phi_{27}$})+\left|t_3\right|  \left|t_6\right|  \cos (\text{$\phi_{36}$})+\left|t_4\right|  \left|t_5\right|  \cos (\text{$\phi_{45}$}) & \ev{\Gamma^{\text{VIII}}_{\text{c1}}}{t} & -\mathcal{O}_{\text{zz'}}^{\text{c}} \\
    \mathcal{O}^{\text{VIII}}_{\text{c2}} & \left|t_1\right|  \left|t_8\right|  \cos (\text{$\phi_{18}$})+\left|t_2\right|  \left|t_7\right|  \cos (\text{$\phi_{27}$})-\left|t_3\right|  \left|t_6\right|  \cos (\text{$\phi_{36}$})-\left|t_4\right|  \left|t_5\right|  \cos (\text{$\phi_{45}$}) & \ev{\Gamma^{\text{VIII}}_{\text{c2}}}{t} & \mathcal{O}_{\text{xz'}}^{\text{s}} \\
    \mathcal{O}^{\text{VIII}}_{\text{c3}} & \left|t_1\right|  \left|t_8\right|  \cos (\text{$\phi_{18}$})-\left|t_2\right|  \left|t_7\right|  \cos (\text{$\phi_{27}$})+\left|t_3\right|  \left|t_6\right|  \cos (\text{$\phi_{36}$})-\left|t_4\right|  \left|t_5\right|  \cos (\text{$\phi_{45}$}) & \ev{\Gamma^{\text{VIII}}_{\text{c3}}}{t} & -\mathcal{O}_{\text{zx'}}^{\text{s}} \\
    \mathcal{O}^{\text{VIII}}_{\text{c4}} & \left|t_1\right|  \left|t_8\right|  \cos (\text{$\phi_{18}$})-\left|t_2\right|  \left|t_7\right|  \cos (\text{$\phi_{27}$})-\left|t_3\right|  \left|t_6\right|  \cos (\text{$\phi_{36}$})+\left|t_4\right|  \left|t_5\right|  \cos (\text{$\phi_{45}$}) & \ev{\Gamma^{\text{VIII}}_{\text{c4}}}{t} & -\mathcal{O}_{\text{xx'}}^{\text{c}} \\
    \mathcal{O}^{\text{VIII}}_{\text{s1}} & \left|t_1\right|  \left|t_8\right|  \sin (\text{$\phi_{18}$})+\left|t_2\right|  \left|t_7\right|  \sin (\text{$\phi_{27}$})+\left|t_3\right|  \left|t_6\right|  \sin (\text{$\phi_{36}$})+\left|t_4\right|  \left|t_5\right|  \sin (\text{$\phi_{45}$}) & \ev{\Gamma^{\text{VIII}}_{\text{s1}}}{t} & -\mathcal{O}_{\text{zz'}}^{\text{s}} \\
    \mathcal{O}^{\text{VIII}}_{\text{s2}} & \left|t_1\right|  \left|t_8\right|  \sin (\text{$\phi_{18}$})+\left|t_2\right|  \left|t_7\right|  \sin (\text{$\phi_{27}$})-\left|t_3\right|  \left|t_6\right|  \sin (\text{$\phi_{36}$})-\left|t_4\right|  \left|t_5\right|  \sin (\text{$\phi_{45}$}) & \ev{\Gamma^{\text{VIII}}_{\text{s2}}}{t} & -\mathcal{O}_{\text{xz'}}^{\text{c}} \\
    \mathcal{O}^{\text{VIII}}_{\text{s3}} & \left|t_1\right|  \left|t_8\right|  \sin (\text{$\phi_{18}$})-\left|t_2\right|  \left|t_7\right|  \sin (\text{$\phi_{27}$})+\left|t_3\right|  \left|t_6\right|  \sin (\text{$\phi_{36}$})-\left|t_4\right|  \left|t_5\right|  \sin (\text{$\phi_{45}$}) & \ev{\Gamma^{\text{VIII}}_{\text{s3}}}{t} & \mathcal{O}_{\text{zx'}}^{\text{c}} \\
    \mathcal{O}^{\text{VIII}}_{\text{s4}} & \left|t_1\right|  \left|t_8\right|  \sin (\text{$\phi_{18}$})-\left|t_2\right|  \left|t_7\right|  \sin (\text{$\phi_{27}$})-\left|t_3\right|  \left|t_6\right|  \sin (\text{$\phi_{36}$})+\left|t_4\right|  \left|t_5\right|  \sin (\text{$\phi_{45}$}) & \ev{\Gamma^{\text{VIII}}_{\text{s4}}}{t} & -\mathcal{O}_{\text{xx'}}^{\text{s}}
\end{tabular}
\end{ruledtabular}
\end{table*}

\clearpage
\section{Database for two pseudoscalar meson photoproduction}\label{sec:database}
As already mentioned, 64 observables can be measured for two pseudoscalar meson photoproduction using the full three-body kinematics of the reaction. These observables can be organized into three groups: single, double and triple polarization observables, which require either the use of a polarized beam ($\mathcal{B}$) or a polarized target ($\mathcal{T}$) or a recoil polarimeter ($\mathcal{R}$) or a combination of the three.
\Cref{tab:ObsMeasurement} gives an overview of all the observables of each category.
In addition to the unpolarized cross section $\text{I}_0$, there are three observables in each single polarization observable category ($\mathcal{B}$,$\mathcal{T}$,$\mathcal{R}$), nine in each double polarization observable category ($\mathcal{BT}$, $\mathcal{BR}$ and $\mathcal{TR}$) and 27 observables in the triple polarization observable category ($\mathcal{BTR}$). \\
The description of the full three-body kinematics requires five independent variables \cite{Seifen2020}.
In this context, two planes, the reaction plane and the decay plane, are often used \cite{gutz2014,Seifen2020}.
While the reaction plane is defined by the incoming photon and one of the outgoing particles, the decay plane is spanned by the other two outgoing particles.
The angle between the reaction and the decay plane is called $\phi^*$. Integrating over $\phi^*$ makes it possible to treat the three-body final state as a two-body final state, resulting in a reduced number of observables.
In this case, the observables correspond to observables known from single meson photoproduction \cite{BDS}, e.g. the category $\mathcal{B}$ reduces to $\text{I}^\text{c}=\Sigma$, the category $\mathcal{T}$ to $\text{P}_{\text{y}}=T$, the category $\mathcal{R}$ to $\text{P}_{\text{y'}}=P$ (this observable can be also measured as a double polarization observable $-\text{P}_\text{y}^{\text{c}}$ \cite{RobertsOed}) and the category $\mathcal{BT}$ to $\text{P}_\text{x}^\text{s}=H$, $\text{P}_\text{z}^\text{s}=G$, $\text{P}_\text{x}^\odot=F$ and $\text{P}_\text{z}^\odot=E$ \cite{RobertsOed}.

In the case of single pseudoscalar meson photoproduction, quite a lot measurements were performed to determine single and double polarization observables.
An extensive overview over the performed measurements, on the basis of the SAID database \cite{SAID_database}, was brought together recently by Ireland, Pasyuk and Strakovsky \cite{IRELAND2020103752}.

A similar database does not exist yet for double pseudoscalar meson photoproduction.
Thus, for the first time, an extensive overview of measurements of polarization observables for double pseudoscalar meson photoproduction is presented in \cref{tab:collectionMeasurements1,tab:collectionMeasurements2}.

By far the most measurements where performed for the reaction $\gamma p \rightarrow p \pi^0 \pi^0$, as the reaction has the lowest amount of non-resonant background amplitude contributions compared to other isospin-channels \cite{Sokhoyan}.
The most common observable is the unpolarized cross section $\text{I}_0$, followed by the beam asymmetries $\text{I}^\text{s},\text{I}^\text{c}$ and $\text{I}^\odot$.
Even a few double polarization observables in quasi two-body kinematics were measured, i.e. $E$ and $H$.
Until now, no triple polarization observables were extracted, as it is experimentally challenging to measure the polarization of a recoiling particle \cite{Sikora_2009,Brinkmann2017} in addition to a polarized beam and a polarized target.

\begin{table}[h]
\caption{The 64 observables are grouped into eight categories according to the polarization needed to measure these observables (beam ($\mathcal{B}$), target ($\mathcal{T}$) and recoil ($\mathcal{R}$)). The notation used in the original paper of Roberts and Oed \cite{RobertsOed} is used for the observables. The observable $\text{I}_0$ corresponds to the unpolarized cross section.}
\label{tab:ObsMeasurement}
\begin{ruledtabular}
\begin{tabular}{CCL}
    \text{Category} & \text{Subcategory} & \text{Observables} \\
    \midrule
   &  & \text{I}_{0}\\
 \midrule
\multirow{2}{*}{$\mathcal{B}$} & \mathcal{B}_\text{l} & \text{I}^\text{s}, \text{I}^\text{c}\\
& \mathcal{B}_{\odot} & \text{I}^{\odot}\\
\midrule
\mathcal{T} & & \text{P}_\text{x}, \text{P}_\text{y}, \text{P}_\text{z}\\
\midrule
\mathcal{R} & & \text{P}_\text{x'}, \text{P}_\text{y'}, \text{P}_\text{z'}\\
\hline
\hline
\multirow{2}{*}{$\mathcal{BT}$} & \mathcal{B}_\text{l} \mathcal{T}& \text{P}^\text{s}_\text{x}, \text{P}^\text{s}_\text{y}, \text{P}^\text{s}_\text{z}, \text{P}^\text{c}_\text{x}, \text{P}^\text{c}_\text{y}, \text{P}^\text{c}_\text{z}\\
& \mathcal{B}_{\odot} \mathcal{T}& \text{P}^{\odot}_\text{x}, \text{P}^{\odot}_\text{y}, \text{P}^{\odot}_\text{z}\\
\midrule
\multirow{2}{*}{$\mathcal{BR}$} & \mathcal{B}_\text{l} \mathcal{R} &  \text{P}^\text{s}_\text{x'}, \text{P}^\text{s}_\text{y'}, \text{P}^\text{s}_\text{z'}, \text{P}^\text{c}_\text{x'}, \text{P}^\text{c}_\text{y'}, \text{P}^\text{c}_\text{z'}\\
& \mathcal{B}_{\odot} \mathcal{R} & \text{P}^{\odot}_\text{x'}, \text{P}^{\odot}_\text{y'}, \text{P}^{\odot}_\text{z'}\\
\midrule
\multirow{2}{*}{$\mathcal{TR}$} & & \mathcal{O}_\text{xx'}, \mathcal{O}_\text{xy'},\mathcal{O}_\text{xz'},\mathcal{O}_\text{yx'},\mathcal{O}_\text{yy'},\mathcal{O}_\text{yz'},\mathcal{O}_\text{zx'}\\
 & & \mathcal{O}_\text{zy'},\mathcal{O}_\text{zz'}\\
\hline
\hline
\multirow{6}{*}{$\mathcal{BTR}$} & \mathcal{B}_\text{l} \mathcal{TR} & \mathcal{O}_\text{xx'}^\text{s}, \mathcal{O}_\text{xy'}^\text{s},\mathcal{O}_\text{xz'}^\text{s},\mathcal{O}_\text{yx'}^\text{s},\mathcal{O}_\text{yy'}^\text{s},\mathcal{O}_\text{yz'}^\text{s},\mathcal{O}_\text{zx'}^\text{s}\\
 &  & \mathcal{O}_\text{zy'}^\text{s},\mathcal{O}_\text{zz'}^\text{s}\\
  & & \mathcal{O}_\text{xx'}^\text{c}, \mathcal{O}_\text{xy'}^\text{c},\mathcal{O}_\text{xz'}^\text{c},\mathcal{O}_\text{yx'}^\text{c},\mathcal{O}_\text{yy'}^\text{c},\mathcal{O}_\text{yz'}^\text{c},\mathcal{O}_\text{zx'}^\text{c}\\
 &  & \mathcal{O}_\text{zy'}^\text{c},\mathcal{O}_\text{zz'}^\text{c}\\
  & \mathcal{B}_{\odot} \mathcal{TR} & \mathcal{O}_\text{xx'}^{\odot}, \mathcal{O}_\text{xy'}^{\odot},\mathcal{O}_\text{xz'}^{\odot},\mathcal{O}_\text{yx'}^{\odot},\mathcal{O}_\text{yy'}^{\odot},\mathcal{O}_\text{yz'}^{\odot},\mathcal{O}_{zx'}^{\odot}\\
 &  & \mathcal{O}_\text{zy'}^{\odot},\mathcal{O}_\text{zz'}^{\odot}
\end{tabular}
\end{ruledtabular}
\end{table}

\begin{table*}
\caption{A collection of polarization observable measurements for two pseudoscalar meson photoproduction.\label{tab:collectionMeasurements1}}
\begin{ruledtabular}
\begin{tabular}{clll}
Observable & Energy range $E^{lab}_\gamma$ & Facility & Reference\\
\midrule
\multicolumn{4}{c}{$\gamma p \rightarrow p \pi^0 \pi^0$}\\
\midrule
$\text{I}_0$ & $309-792$ MeV & TAPS at MAMI & Härter et al. \cite{harter1997}\\
$\text{I}_0$ & $309-820$ MeV & TAPS at MAMI & Wolf et al. \cite{wolf2000}\\
$\text{I}_0$ & $200-820$ MeV & TAPS at MAMI & Kleber et al. \cite{kleber2000}\\
$\text{I}_0$ & $300-425$ MeV & TAPS at MAMI & Kotulla et al. \cite{kotulla2004}\\
$\text{I}_0$ & $309-800$ MeV & CB/TAPS at MAMI &Zehr et al. \cite{zehr2012}\\
$\text{I}_0$ & $309-1400$ MeV & CB/TAPS at MAMI & Kashevarov et al. \cite{kashevarov2012}\\
$\text{I}_0$ & $432-1374$ MeV & CB/TAPS at MAMI & Dieterle et al. \cite{dieterle2015}\\
$\text{I}_0$ & $400-800$ MeV & DAPHNE at MAMI & Braghieri et al. \cite{braghieri1995}\\
$\text{I}_0$ & $400-800$ MeV & DAPHNE at MAMI & Ahrens et al. \cite{ahrens2005}\\
$\text{I}_0$ & $309-820$ MeV & A2+TAPS at MAMI, CB-ELSA & Sarantsev et al. \cite{sarantsev2008}\\
$\text{I}_0$ & $400-1300$ MeV & CB-ELSA & Thoma et al. \cite{thoma2008}\\
$\text{I}_0$ & $\sim 750-2500$ MeV & CB-ELSA/TAPS at ELSA & Thiel et al. \cite{thiel2015}\\
$\text{I}_0, \Sigma$ & $600-2500$ MeV & CB-ELSA/TAPS at ELSA & Sokhoyan et al. \cite{sokhoyan2015}\\
$\text{I}_0$ & $650-1500$ MeV & GRAAL & Assafiri et al. \cite{assafiri2003}\\
$\Sigma$ & $650-1450$ MeV & GRAAL & Assafiri et al. \cite{assafiri2003}\\
$\Sigma$ & $650-1450$ MeV & CB-ELSA & Thoma et al. \cite{thoma2008}\\
$\text{I}^\odot$ & $560-810$ MeV & CB/TAPS at MAMI & Krambrich et al. \cite{krambrich2009}\\
$\text{I}^\odot$ & $\sim 600-1400$ MeV & CB/TAPS at MAMI & Oberle et al. \cite{oberle2013}\\
$\text{I}^\odot$ & $550-820$ MeV & CB/TAPS at MAMI & Zehr et al. \cite{zehr2012}\\
$E, \sigma_{1/2}, \sigma_{3/2}$ & $\sim 431-1455$ MeV & CB/TAPS at MAMI & Dieterle et al. \cite{dieterle2020}\\
$\text{P}_\text{x}, \text{P}_\text{y}, T, H, P$ & $650 - 2600$ MeV & CB-ELSA/TAPS at ELSA & Seifen et al. \cite{Seifen2020}\\
$\text{I}^\text{c}, \text{I}^\text{s}$ & $970-1650$ MeV & CB-ELSA/TAPS at ELSA & Sokhoyan et al. \cite{sokhoyan2015}\\
\midrule
\multicolumn{4}{c}{$\gamma p \rightarrow p \pi^+ \pi^-$}\\
\midrule
$\text{I}_0$ & $400-800$ MeV & DAPHNE at MAMI & Braghieri et al. \cite{braghieri1995}\\
$\text{I}_0$ & $400-800$ MeV & DAPHNE at MAMI & Ahrens et al. \cite{ahrens2007}\\
$\text{I}_0$ & $370-940$ MeV & LNF & Carbonara et al. \cite{carbonara1976}\\
$\text{I}_0$ & $800-1100$ MeV & NKS at LNS & Hirose et al. \cite{hirose2009}\\
$\text{I}_0$ & $500-4800$ MeV & CEA & Crouch et al. \cite{crouch1964}\\
$\text{I}_0$ & $\sim 560 - 2560$ MeV & SAPHIR at ELSA & Wu et al. \cite{wu2005}\\
$\text{I}^\odot$ & $575 - 815$ MeV & TAPS at MAMI & Krambrich et al. \cite{krambrich2009}\\
$\text{I}^\odot$ & $502-2350$ MeV & CLAS at JLAB & Strauch et al. \cite{strauch2005}\\
$\text{I}^\odot$ & $1100-5400$ MeV & CLAS at JLAB & Badui et al. \cite{badui2016}
\end{tabular}
\end{ruledtabular}
\end{table*}
\begin{table*}
\caption{(Continuation of \cref{tab:collectionMeasurements1}) A collection of polarization observable measurements for two pseudoscalar meson photoproduction.\label{tab:collectionMeasurements2}}
\begin{ruledtabular}
\begin{tabular}{clll}
Observable & Energy range $E^{lab}_\gamma$ & Facility & Reference\\
\midrule
\multicolumn{4}{c}{$\gamma p \rightarrow p \pi^0 \eta$}\\
\midrule
$\text{I}_0$ & $\sim 930-2500$ MeV & CBELSA/TAPS at ELSA & Gutz et al. \cite{gutz2014}\\
$\text{I}_0$ & $\sim 1070-2860$ MeV & CBELSA at ELSA & Horn et al. \cite{horn2008}\\
$\text{I}_0$ & $950-1400$ MeV & CB/TAPS at MAMI & Kashevarov et al. \cite{kashevarov2009}\\
$\text{I}_0$ & $1000-1150$ MeV & GeV-$\gamma$ at LNS & Nakabayashi et al. \cite{nakabayashi2006}\\
$\text{I}_0, \Sigma$ & $\sim 930-1500$ MeV & GRAAL & Ajaka et al. \cite{ajaka2008}\\
$\Sigma$ & $970-1650$ MeV & CBELSA/TAPS at ELSA & Gutz et al. \cite{gutz2008}\\
$\Sigma$ & $\sim1070-1550$ MeV & CBELSA/TAPS at ELSA & Gutz et al. \cite{gutz2014}\\
$\text{I}^\text{c}, \text{I}^\text{s}$ & $970-1650$ MeV & CBELSA/TAPS at ELSA & Gutz et al. \cite{gutz2010}\\
$\text{I}^\text{c}, \text{I}^\text{s}$ & $\sim1081-1550$ MeV & CBELSA/TAPS at ELSA & Gutz et al. \cite{gutz2014}\\
\midrule
\multicolumn{4}{c}{$\gamma p \rightarrow n \pi^+ \pi^0$}\\
\midrule
$\text{I}_0$ & $300-820$ MeV & TAPS at MAMI & Langgärtner et al. \cite{langgartner2001}\\
$\text{I}_0$ & $\sim 325-800$ MeV & CB/TAPS at MAMI & Zehr et al. \cite{zehr2012}\\
$\text{I}_0$ & $400-800$ MeV & DAPHNE at MAMI & Braghieri et al. \cite{braghieri1995}\\
$\text{I}_0$ & $400-800$ MeV & DAPHNE at MAMI & Ahrens et al. \cite{ahrens2003}\\
$\text{I}^\odot$ & $520 - 820$ MeV & CB/TAPS at MAMI & Krambrich et al. \cite{krambrich2009}\\
$\text{I}^\odot$ & $\sim 550-820$ MeV & CB/TAPS at MAMI & Zehr et al. \cite{zehr2012}\\
\midrule
\multicolumn{4}{c}{$\gamma n \rightarrow n \pi^0 \pi^0$}\\
\midrule
$\text{I}_0, \Sigma$ & $\sim 600-1500$ MeV & GRAAL & Ajaka et al. \cite{ajaka2007}\\
$\text{I}_0$ & $\sim 430-1371$ MeV & CB/TAPS at MAMI & Dieterle et al. \cite{dieterle2015}\\
\midrule
\multicolumn{4}{c}{$\gamma n \rightarrow p \pi^- \pi^0$}\\
\midrule
$\text{I}_0$ & $\sim370-940$ MeV & LNF & Carbonara et al. \cite{carbonara1976}\\
$\text{I}_0$ & $\sim 450-800$ MeV & DAPHNE at MAMI & Zabrodin et al. \cite{zabrodin1997}\\
$\text{I}_0$ & $\sim 500-800$ MeV & DAPHNE at MAMI & Zabrodin et al. \cite{zabrodin1999}\\
\midrule
\multicolumn{4}{c}{$\gamma n \rightarrow n \pi^+ \pi^-$}\\
\midrule
$\text{I}_0$ & $370-940$ MeV & LNF & Carbonara et al. \cite{carbonara1976}\\
\midrule
\multicolumn{4}{c}{$\gamma p \rightarrow p K^+ K^-$}\\
\midrule
$\text{I}^\odot$ & $1100-5400$ MeV & CLAS at JLAB & Badui et al. \cite{badui2016}
\end{tabular}
\end{ruledtabular}
\end{table*}

\clearpage
\section{Approach\label{sec:approach}}
\begin{figure}[t]
    \includegraphics[scale=0.4]{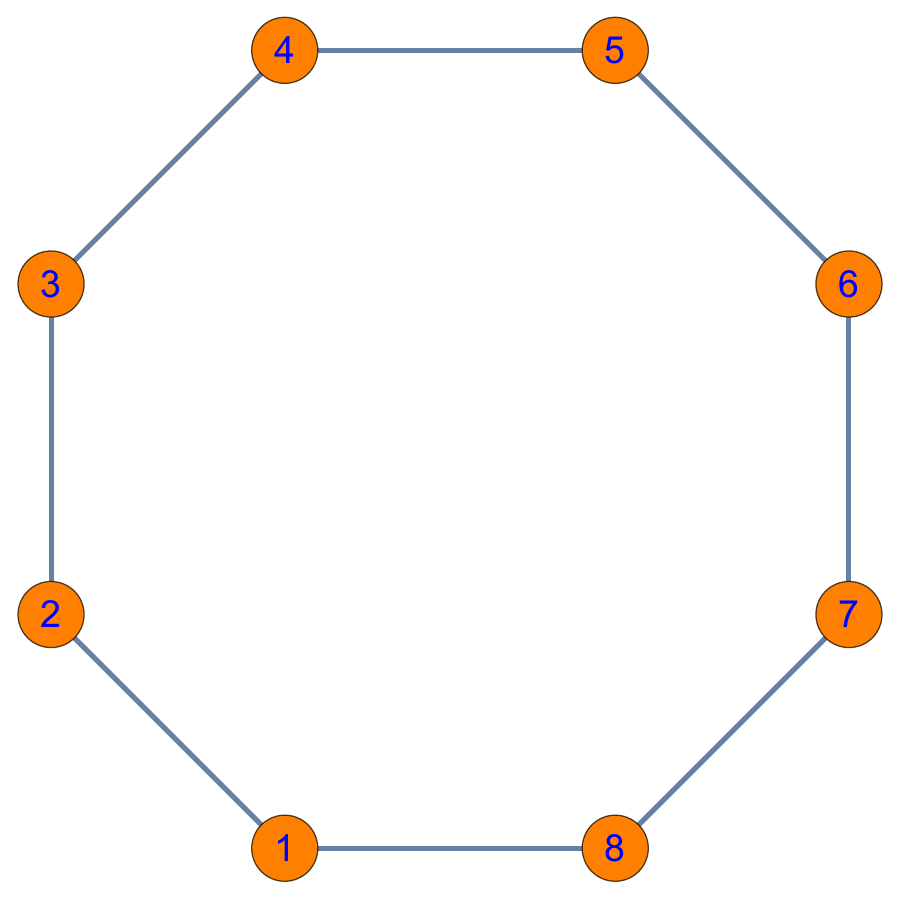}
    \includegraphics[scale=0.4]{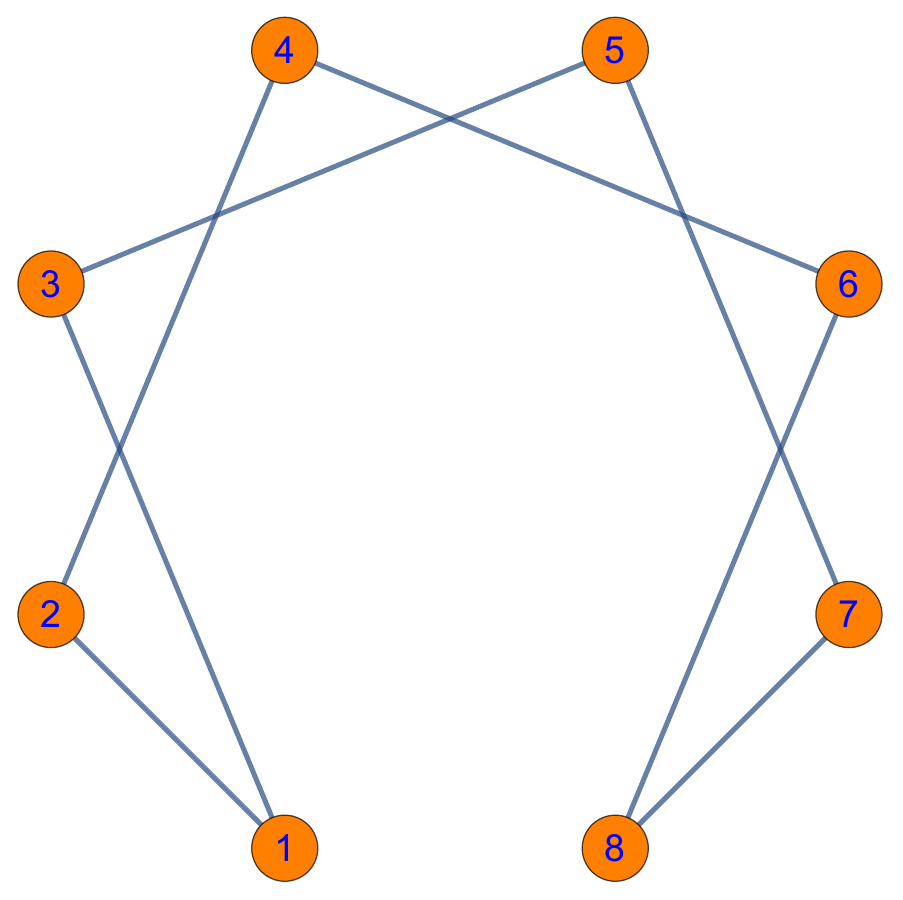}
    \includegraphics[scale=0.4]{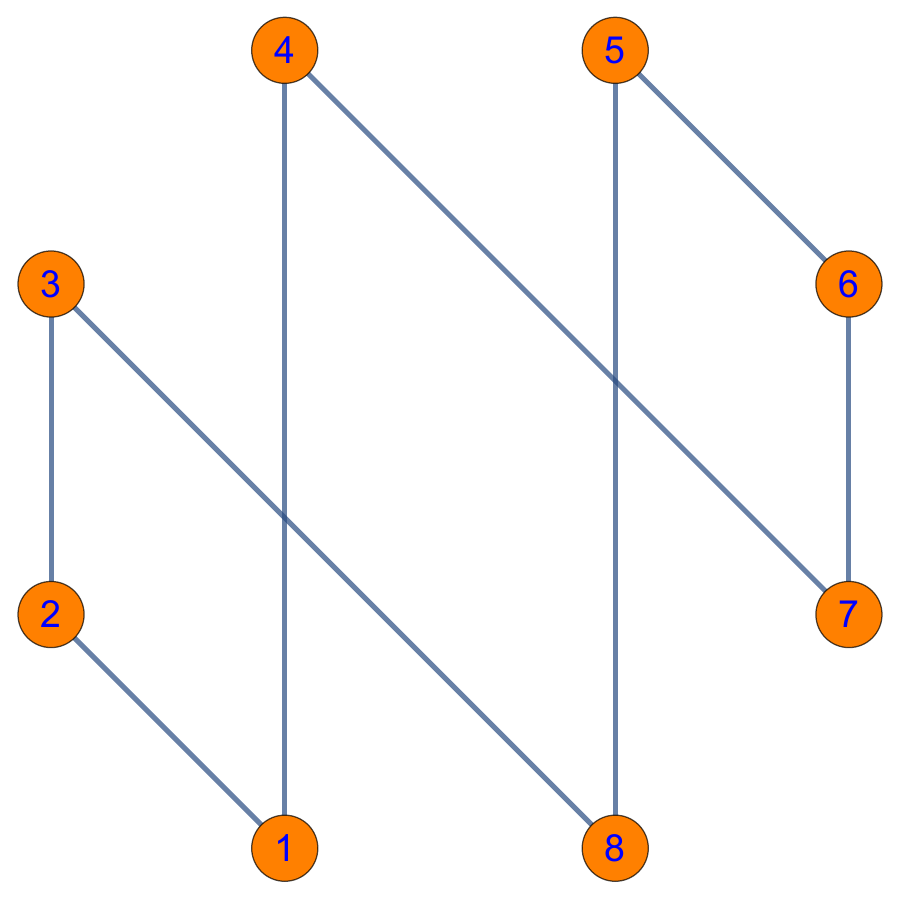}
    \caption{Examples of graph topologies. Only three out of 2520 unique cycle graphs with 8 nodes are shown. \label{graph_example_topologies}}
\end{figure}

The results of \cref{sec:M_theorem} imply the following steps:
One constructs all unique graph topologies with $N$ nodes using combinatorial methods.
A few examples of possible graphs are shown in \cref{graph_example_topologies}.
The number of unique topologies is solely determined by the number of nodes (or edges), i.e. for $N\ge3$ it is $N!/(2N)$ \cite{otherPaperOfUs}.

In a second step, all possible edge configurations which yield a complete set of observables are constructed.
An example is shown in \cref{fig:graph_example}.
This is done for each unique topology.
The total number of possible edge configurations can be calculated by $\sum_{k=1}^N \binom{N}{k}$ for all odd $k\le N$.

The final step involves the mapping from bilinear forms to the actual observables.
Referring again to the example in \cref{fig:graph_example}, the overall question is: Which combinations of observables can be solely described by these bilinear products (given that all amplitudes are known)?
Considering \cref{observables1,observables2}, the following relations are evident:
\begin{align}
    \sim\sin(\phi_{12}) &\mapsto \{\mathcal{O}^\text{III}_{\text{s1}},\mathcal{O}^\text{III}_{\text{s2}},\mathcal{O}^\text{III}_{\text{s3}},\mathcal{O}^\text{III}_{\text{s4}}\},\\
    \sim\sin(\phi_{34}) &\mapsto \{\mathcal{O}^\text{III}_{\text{s1}},\mathcal{O}^\text{III}_{\text{s2}},\mathcal{O}^\text{III}_{\text{s3}},\mathcal{O}^\text{III}_{\text{s4}}\},\\
    \sim\sin(\phi_{56}) &\mapsto \{\mathcal{O}^\text{III}_{\text{s1}},\mathcal{O}^\text{III}_{\text{s2}},\mathcal{O}^\text{III}_{\text{s3}},\mathcal{O}^\text{III}_{\text{s4}}\},\\
    \sim\sin(\phi_{78}) &\mapsto \{\mathcal{O}^\text{III}_{\text{s1}},\mathcal{O}^\text{III}_{\text{s2}},\mathcal{O}^\text{III}_{\text{s3}},\mathcal{O}^\text{III}_{\text{s4}}\}\\
    \sim\sin(\phi_{14}) &\mapsto \{\mathcal{O}^\text{IV}_{\text{s1}},\mathcal{O}^\text{IV}_{\text{s2}},\mathcal{O}^\text{IV}_{\text{s3}},\mathcal{O}^\text{IV}_{\text{s4}}\},\\
    \sim\sin(\phi_{58}) &\mapsto \{\mathcal{O}^\text{IV}_{\text{s1}},\mathcal{O}^\text{IV}_{\text{s2}},\mathcal{O}^\text{IV}_{\text{s3}},\mathcal{O}^\text{IV}_{\text{s4}}\},\\
    \sim\sin(\phi_{27}) &\mapsto \{\mathcal{O}^\text{VIII}_{\text{s1}},\mathcal{O}^\text{VIII}_{\text{s2}},\mathcal{O}^\text{VIII}_{\text{s3}},\mathcal{O}^\text{VIII}_{\text{s4}}\},\\
    \notag\\
    \sim\cos(\phi_{36}) &\mapsto \{\mathcal{O}^\text{VIII}_{\text{c1}},\mathcal{O}^\text{VIII}_{\text{c2}},\mathcal{O}^\text{VIII}_{\text{c3}},\mathcal{O}^\text{VIII}_{\text{c4}}\}.
\end{align}
Thus, the complete set of observables which corresponds to the graph configuration shown in \cref{fig:graph_example} is:
\begin{align}
    \{&\mathcal{O}^\text{I}_1, \mathcal{O}^\text{I}_2, \mathcal{O}^\text{I}_3, \mathcal{O}^\text{I}_4, \mathcal{O}^\text{I}_5, \mathcal{O}^\text{I}_6, \mathcal{O}^\text{I}_7, \mathcal{O}^\text{I}_8,\notag\\
    &\mathcal{O}^\text{III}_{\text{s1}}, \mathcal{O}^\text{III}_{\text{s2}},\mathcal{O}^\text{III}_{\text{s3}}, \mathcal{O}^\text{III}_{\text{s4}}, \mathcal{O}^\text{IV}_{\text{s1}}, \mathcal{O}^\text{IV}_{\text{s2}},\mathcal{O}^\text{IV}_{\text{s3}}, \mathcal{O}^\text{IV}_{\text{s4}}, \label{eq:complete set example}\\
    &\mathcal{O}^\text{VIII}_{\text{s1}}, \mathcal{O}^\text{VIII}_{\text{s2}}, \mathcal{O}^\text{VIII}_{\text{s3}}, \mathcal{O}^\text{VIII}_{\text{s4}}, \mathcal{O}^\text{VIII}_{\text{c1}}, \mathcal{O}^\text{VIII}_{\text{c2}},\mathcal{O}^\text{VIII}_{\text{c3}}, \mathcal{O}^\text{VIII}_{\text{c4}}\}\notag.
\end{align}
In general one needs to add the observables which are solely described by moduli, in order to fix the moduli of the complex amplitudes $t_i$ (see \cref{sec:M_theorem}).
In this case, these are the group I observables, as shown in \cref{observables1}.
Thus, \cref{eq:complete set example} accounts to a total of 24 observables.

The same result can be obtained by using the relation:
\begin{align}
t^*_j t_i = \frac{1}{8} \sum^{64}_{\alpha=1} \Gamma^\alpha_{ij} \mathcal{O}^\alpha, \label{eq::shortForm}
\end{align}
where $\alpha$ is an index running through the observables listed in \cref{observables1,observables2} and the $\Gamma-$matrices as listed in \cref{definition_gamma_matrices}.
\Cref{eq::shortForm} is derived by using $\mathcal{O}^\alpha = \sum^8_{i,j=1} t^*_i \Gamma^\alpha_{ij} t_j$ in combination with the completeness relation of the $\Gamma$-matrices $\sum^{64}_{\alpha=1} \Gamma^\alpha_{ai}\Gamma^\alpha_{jb} = 8\cdot\delta_{ab}\delta_{ij}$.

\section{Results for N=8\label{sec:results}}
For $N=8$ one has 2520 unique cycle graphs, each with 128 unique edge configurations, as explained in \cref{sec:approach}.
Hence, there exist in total $128\cdot2520=322560$ edge configurations which yield a complete set of observables.
However, the resulting sets are not all linearly independent.

The whole algorithm was implemented in MATHEMATICA \cite{Mathematica}, but can just as easily be implemented in other languages such as JULIA \cite{bezanson2017}.
Filtering out the redundant sets, one is left with $5964$ unique sets of observables.
The length of the sets varies between the topologies as well as between different edge configurations.
To be exact, it varies between a total of 24 and 40 observables.

Without loss of generality, the further analysis focuses on the 392 distinct sets with 24 observables.
A numerical analysis was performed which showed that these sets are indeed complete.
The applied algorithm is described in \cref{algorithm}.

Further characteristics of the minimal sets according to Moravcsik involve:
\begin{itemize}
\item
each set inhibits at least five triple polarization observables,
\item
the sets are constructed from four or five different shape classes.
\end{itemize}
However, these sets are slightly over complete since each observable depends on more than one bilinear product.
According to the current knowledge \cite{otherPaperOfUs,PhysRevC.89.034003} a truly minimal complete set consists of $2N$ observables.
Thus the task remains to reduce the slightly over complete sets by eight observables while retaining the completeness.

\clearpage
\section{Reduction to minimal sets of 2N\label{sec:reduction}}
\subsection{Numerical calculation}\label{sec:reductionNumeric}
The smallest complete sets, which emerge from the modified version of Moravcsik's theorem, have a length of 24 (for $N=8$).
Eight of these observables can not be omitted, namely the group I observables, as discussed in \cref{sec:approach}.
From the remaining 16 observables one constructs all possible subsets containing eight observables, which amounts to $\binom{16}{8}=12870$ distinct sets.

In principle this is done for all sets with length of 24, leading to just over five million minimal, complete set candidates.
This number can be further reduced by $\sim7.7\%$, by noting that sets containing only one or two distinct observable groups (apart from group I) do not correspond to a connected graph and thus do not form a complete set.
There are also a few cases in which three distinct observable groups (apart from group I) are not able to form a connected graph, i.e. \{II,III,IV\}, \{II,V,VI\}, \{II,VII,VIII\}, \{III,V,VII\}, \{III,VI,VIII\}, \{IV,V,VIII\} and \{IV,VI,VII\}.

However, due to the enormous number of possible candidate sets just a minor excerpt was analyzed for this paper.
The sets of interest are checked for completeness via a numerical analysis.
The employed algorithm is described in \cref{algorithm}.

So far 4185 unique truly minimal sets of length $2N=16$ have been found.
There are two major differences to the slightly over complete sets with 24 observables.
On the one hand, all sets found are constructed from exactly four different shape classes.
On the other hand, truly minimal complete sets exist with a minimal number of triple polarization observables, namely only $\mathcal{O}_{4}^{I}=\mathcal{O}_{\text{yy'}}^{\odot}$ from group I.
Hence, this observable has to be included in every set as explained in \cref{sec:M_theorem}.

In total 69 sets with only one triple polarization observable have been found.
All of them are shown in \cref{table:trulyMinimalSetsNoTriplePola}.
Hence, these are the most promising ones when it comes to the experimental verification of Moravcsik's theorem.

\subsection{Algebraic phase-fixing method}\label{sec::algebraic_method}
In the following, the phase-fixing approach first developed by Nakayama in a treatment of single-meson photoproduction (i.e. for $N = 4$ amplitudes) \cite{Nakayama} is adapted to two meson photoproduction.
Thus it is possible, although tedious, to derive minimal complete sets of observables with algebraic methods.

Since the full mathematical derivation is quite extensive, all mathematical details are given in a supplementary note \cite{supplementMaterial}.

One starts by combining, i.e. adding or subtracting, the observables within one shape-class, in such a way that the result only depends on two relative phases.
By doing this, a “decoupled" shape-class is formed.
See also \cref{tab:DecoupledShapeClassesVsRelativePhasesNew}.
\begin{table}
\caption{The $14$ decoupled shape-classes IIa, IIb, $\ldots$, VIIIa, VIIIb are listed together with their corresponding pairs of relative phases.\label{tab:DecoupledShapeClassesVsRelativePhasesNew}}
\begin{ruledtabular}
 \begin{tabular}{rr}
  IIa $\longrightarrow$  $\left\{ \phi_{13}, \phi_{24} \right\}$ & IIb $\longrightarrow$  $\left\{ \phi_{57}, \phi_{68} \right\}$ \\
  IIIa $\longrightarrow$  $\left\{ \phi_{12}, \phi_{34} \right\}$ & IIIb $\longrightarrow$  $\left\{ \phi_{56}, \phi_{78} \right\}$ \\
  IVa $\longrightarrow$  $\left\{ \phi_{14}, \phi_{23} \right\}$ & IVb $\longrightarrow$  $\left\{ \phi_{58}, \phi_{67} \right\}$ \\
  Va $\longrightarrow$  $\left\{ \phi_{15}, \phi_{26} \right\}$ & Vb $\longrightarrow$  $\left\{ \phi_{37}, \phi_{48} \right\}$ \\
  VIa $\longrightarrow$  $\left\{ \phi_{17}, \phi_{28} \right\}$ & VIb $\longrightarrow$  $\left\{ \phi_{35}, \phi_{46} \right\}$ \\
  VIIa $\longrightarrow$  $\left\{ \phi_{16}, \phi_{25} \right\}$ & VIIb $\longrightarrow$  $\left\{ \phi_{38}, \phi_{47} \right\}$ \\
  VIIIa $\longrightarrow$  $\left\{ \phi_{18}, \phi_{27} \right\}$ & VIIIb $\longrightarrow$  $\left\{ \phi_{36}, \phi_{45} \right\}$ \\
 \end{tabular}
\end{ruledtabular}
\end{table}
In that way, one establishes a mathematical similarity with the shape-classes in single-meson photoproduction \cite{Nakayama,otherPaperOfUs}.
For shape-class II this would be:
\begin{align}
&\text{IIa}: \mathcal{O}^\text{II}_\text{s1}+\mathcal{O}^\text{II}_\text{s2},\mathcal{O}^\text{II}_\text{s3}+\mathcal{O}^\text{II}_\text{s4},\mathcal{O}^\text{II}_\text{c1}+\mathcal{O}^\text{II}_\text{c2},\mathcal{O}^\text{II}_\text{c3}+\mathcal{O}^\text{II}_\text{c4},\\
&\text{IIb}: \mathcal{O}^\text{II}_\text{s1}-\mathcal{O}^\text{II}_\text{s2}, \mathcal{O}^\text{II}_\text{s3}-\mathcal{O}^\text{II}_\text{s4}, \mathcal{O}^\text{II}_\text{c1}-\mathcal{O}^\text{II}_\text{c2}, \mathcal{O}^\text{II}_\text{c3}-\mathcal{O}^\text{II}_\text{c4}.
\end{align}

The algebraic approach shown here works out only in case the observables are selected from very particular combinations of four decoupled shape-classes.

More precisely, it has to be possible to establish a “consistency relation" (cf. \cref{eqn:consistency relation}) among the relative phases belonging to all the involved decoupled shape-classes.
A necessary and sufficient condition for this can be formulated in terms of the graph constructed from the relative phases (cf. \cref{sec:M_theorem}): the latter graph has to be a cycle graph.

There exist 40 possible combinations of four decoupled shape classes fulfilling these requirements and which have the following general form:
\begin{align}
\{\text{Xa}, \text{Xb}, \text{Y}, \text{Z}\}. \label{eq:generalFormAlgebraic}
\end{align}
Two examples are shown in \cref{fig:graph_examples_consistencyRelNew}, a complete list can be found in the supplementary material \cite{supplementMaterial}.
The following derivation holds for all combinations of shape-classes of the form given in \cref{eq:generalFormAlgebraic}.
\begin{figure}[t]
    \includegraphics[width=0.22\textwidth]{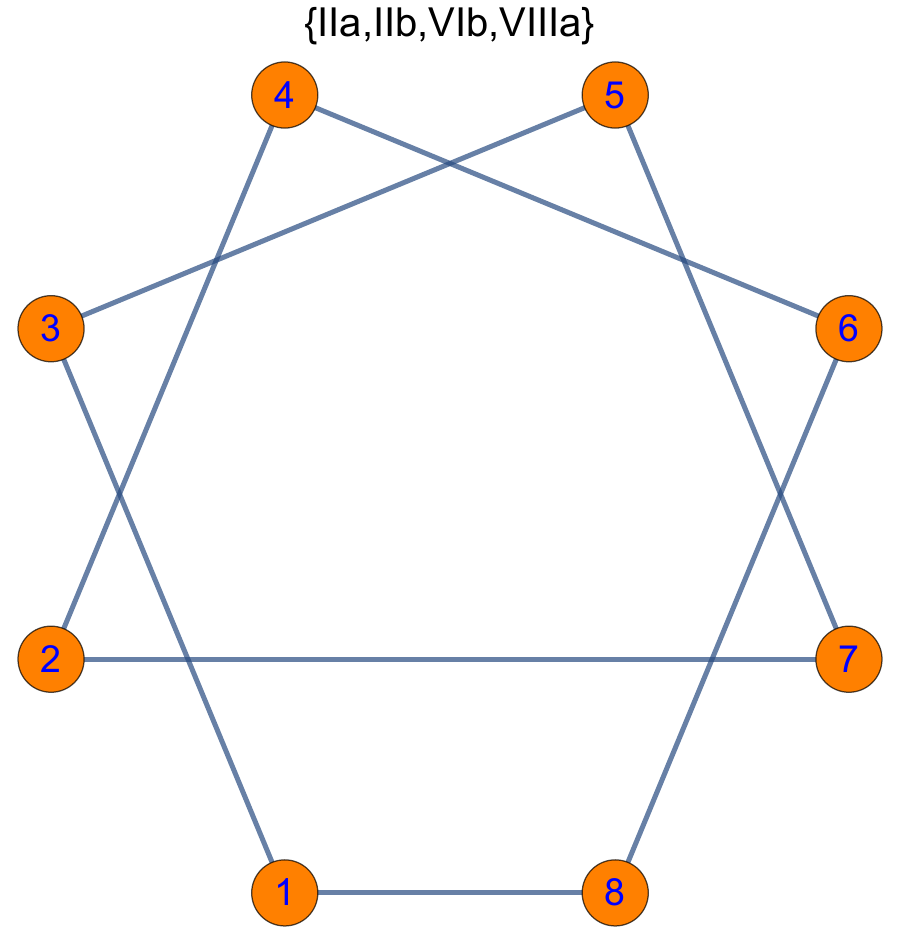} \hspace*{10pt}
    \includegraphics[width=0.22\textwidth]{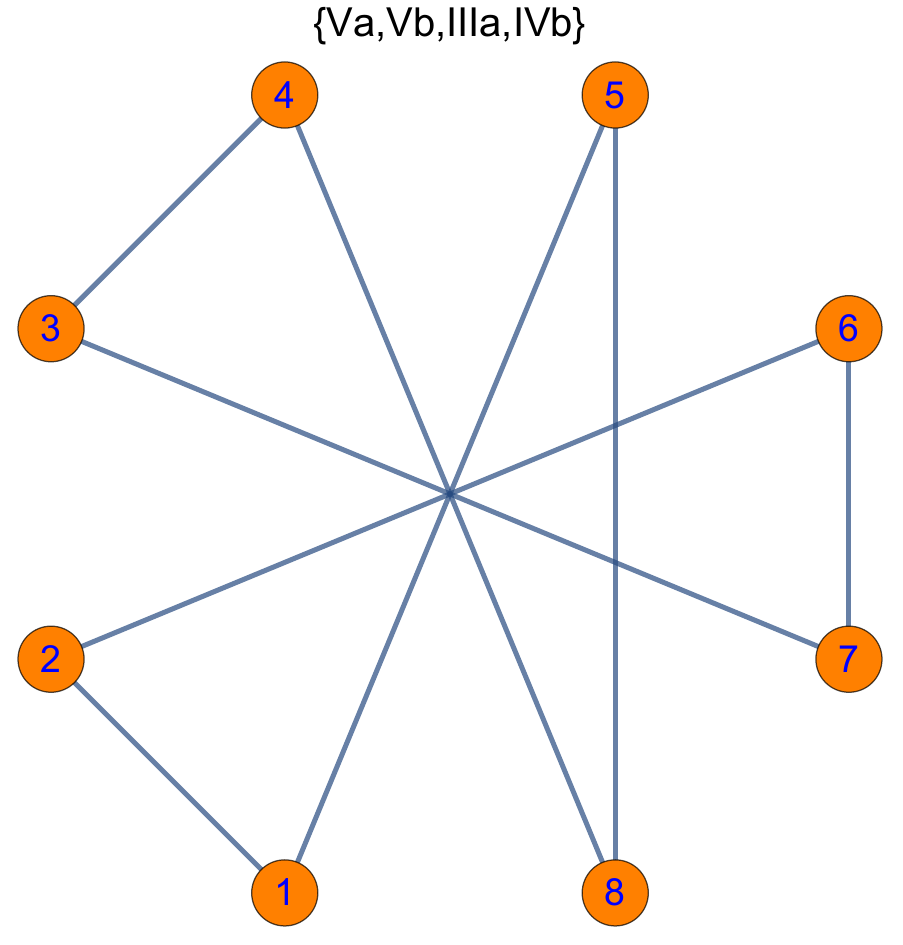}
    \caption{Examples are shown for graph topologies which imply the possibility for a consistency relation (cf. the relative phases listed in \cref{tab:DecoupledShapeClassesVsRelativePhasesNew}). \label{fig:graph_examples_consistencyRelNew}}
\end{figure}

The following discussion focuses on the example case:
\begin{align}
\{\text{IIa}, \text{IIb}, \text{VIb}, \text{VIIIa}\}. \label{eq:NeededShapeClassCombinationExampleNew}
\end{align}
For the remaining 39 cases, the derivation proceeds analogously.

For the example case \cref{eq:NeededShapeClassCombinationExampleNew}, the consistency relation reads (cf. \cref{tab:DecoupledShapeClassesVsRelativePhasesNew}):
\begin{equation}
 \underbrace{\phi_{13} + \phi_{24}}_{\text{IIa}} + \underbrace{\phi_{57} + \phi_{68}}_{\text{IIb}} =  \underbrace{\phi_{18} + \phi_{27}}_{\text{VIIIa}} - \underbrace{\phi_{35} - \phi_{46}}_{\text{VIb}}. \label{eq:ConsistencyRelationExampleNew}
\end{equation}
According to Nakayama \cite{Nakayama}, the “phase-fixing" procedure starts by picking a particular combination of observables from the considered combination of shape-classes, i.e. from \cref{eq:NeededShapeClassCombinationExampleNew}.
In general, one picks two observables from shape-class $\text{Xa}$, two from $\text{Xb}$ and one observable each from two different shape-classes selected from the 12 remaining.
For the case at hand, these are two observables from shape-class IIa, two from IIb, one from VIb as well as one from VIIIa.
For any selection of observables which has this pattern, one has to work out the remaining discrete phase-ambiguities which exist for the associated relative-phases.
For each combination of possible discrete ambiguities, the consistency relation \cref{eq:ConsistencyRelationExampleNew} then has to be evaluated.
In case the consistency relations of such a combination are all linearly independent \cite{Nakayama,otherPaperOfUs}, the considered set of observables is complete.

The way forward is analogous to \cite{Nakayama}.
A detailed description, on how to determine all possible discrete ambiguities and determining whether a set of consistency relations is linear independent or not, can be found in the supplementary material \cite{supplementMaterial}.

The result for the discussed example can be found in \cref{tab:YannicksResults}.
The shown results come only from considering the left side of \cref{eq:ConsistencyRelationExampleNew}.
In general, the determination of the discrete ambiguities of the left side is easier than that of the right side.
So theoretically even more combinations are possible.
Unfortunately, the minimal sets shown in \cref{tab:YannicksResults} contain at least three triple polarization observables.
\vspace{-7pt}
\section{Implications for experimentalists}\label{sec:experimentalImplications}
The experimental verification of complete sets, in the framework of single pseudoscalar meson photoproduction, was studied for example by Ireland \cite{Ireland2010} and Vrancx et al. \cite{Vrancx2013}.
They concluded that as soon as the data have a finite measurement uncertainty, it might not be possible to determine unique solutions for the amplitudes in case one uses a mathematical, minimal complete set of observables.
It is believed that additional obervables could resolve the ambiguity.
Thus it is important to measure more than 16 observables \cite{Ireland2010}.

Observables which are relatively easy to measure rank in the categories $\mathcal{B}, \mathcal{T}, \mathcal{BT}$ as they do not require a recoil polarimeter.
Even though they form a group of 16 observables, they do not form a complete set, which was verified numerically in this study.
The measurement of some well chosen observables from the categories $\mathcal{R}$, $\mathcal{BR}$, $\mathcal{TR}$ and $\mathcal{BTR}$ is essential.
Out of the 69 possible complete sets, that require the measurement of only one triple polarization observable and which are listed in Table \ref{table:trulyMinimalSetsNoTriplePola}, the sets $6)$, $12)$, $15)$, $27)$, $30)$, $41)$ and $47)$ contain the minimal number of recoil polarization observables, e.g. the set 15) contains the following observables:
\begin{multline}
    \{\text{I}^\odot, \text{P}_\text{y}, \text{P}_\text{y'}, \mathcal{O}^\odot_\text{yy’}, \mathcal{O}_\text{yy’}, \text{P}^\odot_\text{y'}, \text{P}^\odot_\text{y}, \text{I}_0,\\
    \text{P}_\text{x}, \text{P}_\text{z}, \text{P}_{\text{x'}}, \text{P}^\text{s}_\text{x}, \text{P}^\odot_\text{x}, \text{P}^\text{c}_\text{z}, \text{P}^\odot_\text{z}, \text{P}^\odot_\text{x'}\}.
\end{multline}
There exist data for 8 of these observables ($\text{I}_0$, $\text{I}^\odot$, $\text{P}_\text{x}$, $\text{P}_\text{y}$, $\text{P}_\text{y'}$, $\mathcal{O}_\text{yy’}(=-\text{I}^\text{c})$, $\text{P}^\text{s}_\text{x}$, $\text{P}^\odot_\text{z}$) for the $p\pi^0\pi^0$ final state, albeit not having a perfect overlap of the energy and angular ranges covered by the different data sets (see \cref{tab:collectionMeasurements1,tab:collectionMeasurements2}). The remaining 8 observables could be measured in the future in three different experiments using a linearly polarized photon beam with a longitudinally polarized target ($\text{P}_{\text{z}}$, $\text{P}_{\text{z}}^\text{c}$), using a circularly polarized photon beam and a transversely polarized target ($\text{P}^\odot_\text{x}$, $\text{P}^\odot_\text{y}$) and by employing a recoil polarimeter in addition to the latter configuration, the observables $\text{P}_{\text{x'}}$, $\text{P}^\odot_{\text{x'}}$, $\text{P}^\odot_{\text{y'}}$ and $\mathcal{O}^\odot_\text{yy’}$ can be obtained as well.
\vspace{-5pt}
\section{Conclusion and Outlook\label{sec:summary}}
Within this paper, the problem of findining complete sets for two pseudoscalar meson photoproduction was studied.
For this purpose, a slightly modified version of Moravcsik's theorem was applied.
This method is capable of extracting complete sets of observables in a totally automated manner.
The automation capability, easy accessibility as well as the adaptability for reactions with arbitrary $N$ are the strengths of Moravcsik's theorem.
However, it turns out that the resulting sets from Moravcsik are slightly over-complete since each observable depends on more than one bilinear product.
For this reason, a numerical- as well as an algebraic method is discussed in order to reduce these sets to minimal complete sets containing $2N=16$ observables.
The characteristics of the minimal sets are discussed.
Finally, 69 minimal complete sets containing the minimal number of triple polarization observables, namely only one, are presented.
From these subsets in combination with the extensive overview of already performed measurements in two pseudoscalar meson photoproduction, the most promising set of observables, which could be measured in the near future, is presented.

Further studies could be performed on how Moravcsik's theorem should be adapted in order to yield directly minimal complete sets.
This would decrease the numerical effort enormously and would make the theorem even more accessible.

\begin{acknowledgments}
The authors would like to thank Prof. Dr. Thoma, for a fruitful discussion and constructive comments on the manuscript and Prof. Dr. Beck for his support.
\end{acknowledgments}

\begin{ruledtabular}
\begin{table*}
\caption{Truly minimal sets consisting of $2N=16$ observables with the minimal number of triple polarization observables, i.e. just $\mathcal{O}_{4}^{I}=\mathcal{O}_{\text{yy'}}^{\odot}$. However, the group I observables are not explicitly shown, as they are common to all complete sets. The used notation is analogous to Roberts/Oed \cite{RobertsOed}.\label{table:trulyMinimalSetsNoTriplePola}}
\begin{minipage}{0.45\textwidth}
\begin{tabular}{CLLLLLLLL}
 1) & \text{P}_{\text{z}} & \text{P}_{\text{x'}} & \text{P}_{\text{z'}} & \text{P}_{\text{x}}^\text{s} & \text{P}_{\text{z}}^\text{c} & \text{P}_{\text{z}}^\odot & \text{P}_{\text{x'}}^\odot & \text{P}_{\text{z'}}^\odot \\
 2) & \text{P}_{\text{z}} & \text{P}_{\text{x}}^\text{s} & \text{P}_{\text{z}}^\text{c} & \text{P}_{\text{z}}^\odot & \text{P}_{\text{x'}}^\odot & \text{P}_{\text{z'}}^\odot & \mathcal{O}_{\text{yx'}} & \mathcal{O}_{\text{yz'}} \\
 3) & \text{P}_{\text{z}} & \text{P}_{\text{x}}^\text{s} & \text{P}_{\text{z}}^\text{c} & \text{P}_{\text{x'}}^\odot & \text{P}_{\text{z'}}^\odot & \mathcal{O}_{\text{yx'}} & \mathcal{O}_{\text{yz'}} & \mathcal{O}_{\text{zy'}} \\
 4) & \text{P}_{\text{x'}} & \text{P}_{\text{z'}} & \text{P}_{\text{x}}^\text{s} & \text{P}_{\text{z}}^\text{c} & \text{P}_{\text{z}}^\odot & \text{P}_{\text{x'}}^\odot & \text{P}_{\text{z'}}^\odot & \mathcal{O}_{\text{zy'}} \\
 5) & \text{P}_{\text{x}}^\text{s} & \text{P}_{\text{z}}^\text{c} & \text{P}_{\text{z}}^\odot & \text{P}_{\text{x'}}^\odot & \text{P}_{\text{z'}}^\odot & \mathcal{O}_{\text{yx'}} & \mathcal{O}_{\text{yz'}} & \mathcal{O}_{\text{zy'}} \\
 6) & \text{P}_{\text{x}} & \text{P}_{\text{z}} & \text{P}_{\text{z'}} & \text{P}_{\text{x}}^\text{s} & \text{P}_{\text{x}}^\odot & \text{P}_{\text{z}}^\text{c} & \text{P}_{\text{z}}^\odot & \text{P}_{\text{z'}}^\odot \\
 7) & \text{P}_{\text{x}} & \text{P}_{\text{z}} & \text{P}_{\text{z'}} & \text{P}_{\text{x}}^\text{s} & \text{P}_{\text{z}}^\text{c} & \text{P}_{\text{z'}}^\odot & \mathcal{O}_{\text{xy'}} & \mathcal{O}_{\text{zy'}}\\
 8) & \text{P}_{\text{x}} & \text{P}_{\text{z}} & \text{P}_{\text{z'}} & \text{P}_{\text{x}}^\text{s} & \text{P}_{\text{z}}^\text{c} & \mathcal{O}_{\text{xy'}} & \mathcal{O}_{\text{yz'}} & \mathcal{O}_{\text{zy'}}\\
 9) & \text{P}_{\text{z'}} & \text{P}_{\text{x}}^\text{s} & \text{P}_{\text{x}}^\odot & \text{P}_{\text{z}}^\text{c} & \text{P}_{\text{z}}^\odot & \text{P}_{\text{z'}}^\odot & \mathcal{O}_{\text{xy'}} & \mathcal{O}_{\text{zy'}}\\
 10) & \text{P}_{\text{z'}} & \text{P}_{\text{x}}^\text{s} & \text{P}_{\text{x}}^\odot & \text{P}_{\text{z}}^\text{c} & \text{P}_{\text{z}}^\odot & \mathcal{O}_{\text{xy'}} & \mathcal{O}_{\text{yz'}} & \mathcal{O}_{\text{zy'}}\\
 11) & \text{P}_{\text{x}}^\text{s} & \text{P}_{\text{x}}^\odot & \text{P}_{\text{z}}^\text{c} & \text{P}_{\text{z}}^\odot & \text{P}_{\text{z'}}^\odot & \mathcal{O}_{\text{xy'}} & \mathcal{O}_{\text{yz'}} & \mathcal{O}_{\text{zy'}}\\
 12) & \text{P}_{\text{z}} & \text{P}_{\text{x}}^\text{c} & \text{P}_{\text{x}}^\text{s} & \text{P}_{\text{z}}^\text{c} & \text{P}_{\text{z}}^\text{s} & \text{P}_{\text{z}}^\odot & \text{P}_{\text{z'}}^\odot & \mathcal{O}_{\text{yz'}}\\
 13) & \text{P}_{\text{z}} & \text{P}_{\text{x}}^\text{c} & \text{P}_{\text{x}}^\text{s} & \text{P}_{\text{z}}^\text{c} & \text{P}_{\text{z}}^\text{s} & \text{P}_{\text{z'}}^\odot & \mathcal{O}_{\text{yz'}} & \mathcal{O}_{\text{zy'}}\\
 14) & \text{P}_{\text{x}}^\text{c} & \text{P}_{\text{x}}^\text{s} & \text{P}_{\text{z}}^\text{c} & \text{P}_{\text{z}}^\text{s} & \text{P}_{\text{z}}^\odot & \text{P}_{\text{z'}}^\odot & \mathcal{O}_{\text{yz'}} & \mathcal{O}_{\text{zy'}}\\
 15) & \text{P}_{\text{x}} & \text{P}_{\text{z}} & \text{P}_{\text{x'}} & \text{P}_{\text{x}}^\text{s} & \text{P}_{\text{x}}^\odot & \text{P}_{\text{z}}^\text{c} & \text{P}_{\text{z}}^\odot & \text{P}_{\text{x'}}^\odot \\
 16) & \text{P}_{\text{x}} & \text{P}_{\text{z}} & \text{P}_{\text{x'}} & \text{P}_{\text{x}}^\text{s} & \text{P}_{\text{z}}^\text{c} & \text{P}_{\text{x'}}^\odot & \mathcal{O}_{\text{xy'}} & \mathcal{O}_{\text{zy'}}\\
 17) & \text{P}_{\text{x}} & \text{P}_{\text{z}} & \text{P}_{\text{x'}} & \text{P}_{\text{x}}^\text{s} & \text{P}_{\text{z}}^\text{c} & \mathcal{O}_{\text{xy'}} & \mathcal{O}_{\text{yx'}} & \mathcal{O}_{\text{zy'}}\\
 18) & \text{P}_{\text{x'}} & \text{P}_{\text{x}}^\text{s} & \text{P}_{\text{x}}^\odot & \text{P}_{\text{z}}^\text{c} & \text{P}_{\text{z}}^\odot & \text{P}_{\text{x'}}^\odot & \mathcal{O}_{\text{xy'}} & \mathcal{O}_{\text{zy'}}\\
 19) & \text{P}_{\text{x'}} & \text{P}_{\text{x}}^\text{s} & \text{P}_{\text{x}}^\odot & \text{P}_{\text{z}}^\text{c} & \text{P}_{\text{z}}^\odot & \mathcal{O}_{\text{xy'}} & \mathcal{O}_{\text{yx'}} & \mathcal{O}_{\text{zy'}}\\
 20) & \text{P}_{\text{x}}^\text{s} & \text{P}_{\text{x}}^\odot & \text{P}_{\text{z}}^\text{c} & \text{P}_{\text{z}}^\odot & \text{P}_{\text{x'}}^\odot & \mathcal{O}_{\text{xy'}} & \mathcal{O}_{\text{yx'}} & \mathcal{O}_{\text{zy'}}\\
 21) & \text{P}_{\text{z}} & \text{P}_{\text{x'}} & \text{P}_{\text{z'}} & \text{P}_{\text{x}}^\text{c} & \text{P}_{\text{z}}^\text{s} & \text{P}_{\text{z}}^\odot & \text{P}_{\text{x'}}^\odot & \text{P}_{\text{z'}}^\odot \\
 22) & \text{P}_{\text{z}} & \text{P}_{\text{x}}^\text{c} & \text{P}_{\text{z}}^\text{s} & \text{P}_{\text{z}}^\odot & \text{P}_{\text{x'}}^\odot & \text{P}_{\text{z'}}^\odot & \mathcal{O}_{\text{yx'}} & \mathcal{O}_{\text{yz'}}\\
 23) & \text{P}_{\text{z}} & \text{P}_{\text{x'}} & \text{P}_{\text{z'}} & \text{P}_{\text{x}}^\text{c} & \text{P}_{\text{z}}^\text{s} & \mathcal{O}_{\text{yx'}} & \mathcal{O}_{\text{yz'}} & \mathcal{O}_{\text{zy'}}\\
 24) & \text{P}_{\text{z}} & \text{P}_{\text{x}}^\text{c} & \text{P}_{\text{z}}^\text{s} & \text{P}_{\text{x'}}^\odot & \text{P}_{\text{z'}}^\odot & \mathcal{O}_{\text{yx'}} & \mathcal{O}_{\text{yz'}} & \mathcal{O}_{\text{zy'}}\\
 25) & \text{P}_{\text{x'}} & \text{P}_{\text{z'}} & \text{P}_{\text{x}}^\text{c} & \text{P}_{\text{z}}^\text{s} & \text{P}_{\text{z}}^\odot & \text{P}_{\text{x'}}^\odot & \text{P}_{\text{z'}}^\odot & \mathcal{O}_{\text{zy'}}\\
 26) & \text{P}_{\text{x}}^\text{c} & \text{P}_{\text{z}}^\text{s} & \text{P}_{\text{z}}^\odot & \text{P}_{\text{x'}}^\odot & \text{P}_{\text{z'}}^\odot & \mathcal{O}_{\text{yx'}} & \mathcal{O}_{\text{yz'}} & \mathcal{O}_{\text{zy'}}\\
 27) & \text{P}_{\text{z}} & \text{P}_{\text{x}}^\text{c} & \text{P}_{\text{x}}^\text{s} & \text{P}_{\text{z}}^\text{c} & \text{P}_{\text{z}}^\text{s} & \text{P}_{\text{z}}^\odot & \text{P}_{\text{x'}}^\odot & \mathcal{O}_{\text{yx'}}\\
 28) & \text{P}_{\text{z}} & \text{P}_{\text{x}}^\text{c} & \text{P}_{\text{x}}^\text{s} & \text{P}_{\text{z}}^\text{c} & \text{P}_{\text{z}}^\text{s} & \text{P}_{\text{x'}}^\odot & \mathcal{O}_{\text{yx'}} & \mathcal{O}_{\text{zy'}}\\
 29) & \text{P}_{\text{x}}^\text{c} & \text{P}_{\text{x}}^\text{s} & \text{P}_{\text{z}}^\text{c} & \text{P}_{\text{z}}^\text{s} & \text{P}_{\text{z}}^\odot & \text{P}_{\text{x'}}^\odot & \mathcal{O}_{\text{yx'}} & \mathcal{O}_{\text{zy'}}\\
 30) & \text{P}_{\text{x}} & \text{P}_{\text{z}} & \text{P}_{\text{z'}} & \text{P}_{\text{x}}^\text{c} & \text{P}_{\text{x}}^\odot & \text{P}_{\text{z}}^\text{s} & \text{P}_{\text{z}}^\odot & \text{P}_{\text{z'}}^\odot \\
 31) & \text{P}_{\text{x}} & \text{P}_{\text{z}} & \text{P}_{\text{z'}} & \text{P}_{\text{x}}^\text{c} & \text{P}_{\text{z}}^\text{s} & \text{P}_{\text{z'}}^\odot & \mathcal{O}_{\text{xy'}} & \mathcal{O}_{\text{zy'}}\\
 32) & \text{P}_{\text{x}} & \text{P}_{\text{z}} & \text{P}_{\text{z'}} & \text{P}_{\text{x}}^\text{c} & \text{P}_{\text{z}}^\text{s} & \mathcal{O}_{\text{xy'}} & \mathcal{O}_{\text{yz'}} & \mathcal{O}_{\text{zy'}}\\
 33) & \text{P}_{\text{z'}} & \text{P}_{\text{x}}^\text{c} & \text{P}_{\text{x}}^\odot & \text{P}_{\text{z}}^\text{s} & \text{P}_{\text{z}}^\odot & \text{P}_{\text{z'}}^\odot & \mathcal{O}_{\text{xy'}} & \mathcal{O}_{\text{zy'}}\\
 34) & \text{P}_{\text{z'}} & \text{P}_{\text{x}}^\text{c} & \text{P}_{\text{x}}^\odot & \text{P}_{\text{z}}^\text{s} & \text{P}_{\text{z}}^\odot & \mathcal{O}_{\text{xy'}} & \mathcal{O}_{\text{yz'}} & \mathcal{O}_{\text{zy'}}\\
 35) & \text{P}_{\text{x}}^\text{c} & \text{P}_{\text{x}}^\odot & \text{P}_{\text{z}}^\text{s} & \text{P}_{\text{z}}^\odot & \text{P}_{\text{z'}}^\odot & \mathcal{O}_{\text{xy'}} & \mathcal{O}_{\text{yz'}} & \mathcal{O}_{\text{zy'}}
\end{tabular}
\end{minipage}\hfill
\begin{minipage}{0.45\textwidth}
\begin{tabular}{CLLLLLLLL}
 36) & \text{P}_{\text{x}} & \text{P}_{\text{x'}} & \text{P}_{\text{z'}} & \text{P}_{\text{x}}^\text{s} & \text{P}_{\text{x}}^\odot & \text{P}_{\text{z}}^\text{c} & \text{P}_{\text{x'}}^\odot & \text{P}_{\text{z'}}^\odot \\
 37) & \text{P}_{\text{x}} & \text{P}_{\text{x}}^\text{s} & \text{P}_{\text{x}}^\odot & \text{P}_{\text{z}}^\text{c} & \text{P}_{\text{x'}}^\odot & \text{P}_{\text{z'}}^\odot & \mathcal{O}_{\text{yx'}} & \mathcal{O}_{\text{yz'}}\\
 38) & \text{P}_{\text{x}} & \text{P}_{\text{x}}^\text{s} & \text{P}_{\text{z}}^\text{c} & \text{P}_{\text{x'}}^\odot & \text{P}_{\text{z'}}^\odot & \mathcal{O}_{\text{xy'}} & \mathcal{O}_{\text{yx'}} & \mathcal{O}_{\text{yz'}}\\
 39) & \text{P}_{\text{x'}} & \text{P}_{\text{z'}} & \text{P}_{\text{x}}^\text{s} & \text{P}_{\text{x}}^\odot & \text{P}_{\text{z}}^\text{c} & \text{P}_{\text{x'}}^\odot & \text{P}_{\text{z'}}^\odot & \mathcal{O}_{\text{xy'}}\\
 40) & \text{P}_{\text{x}}^\text{s} & \text{P}_{\text{x}}^\odot & \text{P}_{\text{z}}^\text{c} & \text{P}_{\text{x'}}^\odot & \text{P}_{\text{z'}}^\odot & \mathcal{O}_{\text{xy'}} & \mathcal{O}_{\text{yx'}} & \mathcal{O}_{\text{yz'}}\\
 41) & \text{P}_{\text{x}} & \text{P}_{\text{z}} & \text{P}_{\text{x'}} & \text{P}_{\text{x}}^\text{c} & \text{P}_{\text{x}}^\odot & \text{P}_{\text{z}}^\text{s} & \text{P}_{\text{z}}^\odot & \text{P}_{\text{x'}}^\odot \\
 42) & \text{P}_{\text{x}} & \text{P}_{\text{z}} & \text{P}_{\text{x'}} & \text{P}_{\text{x}}^\text{c} & \text{P}_{\text{z}}^\text{s} & \text{P}_{\text{x'}}^\odot & \mathcal{O}_{\text{xy'}} & \mathcal{O}_{\text{zy'}}\\
 43) & \text{P}_{\text{x}} & \text{P}_{\text{z}} & \text{P}_{\text{x'}} & \text{P}_{\text{x}}^\text{c} & \text{P}_{\text{z}}^\text{s} & \mathcal{O}_{\text{xy'}} & \mathcal{O}_{\text{yx'}} & \mathcal{O}_{\text{zy'}}\\
 44) & \text{P}_{\text{x'}} & \text{P}_{\text{x}}^\text{c} & \text{P}_{\text{x}}^\odot & \text{P}_{\text{z}}^\text{s} & \text{P}_{\text{z}}^\odot & \text{P}_{\text{x'}}^\odot & \mathcal{O}_{\text{xy'}} & \mathcal{O}_{\text{zy'}}\\
 45) & \text{P}_{\text{x'}} & \text{P}_{\text{x}}^\text{c} & \text{P}_{\text{x}}^\odot & \text{P}_{\text{z}}^\text{s} & \text{P}_{\text{z}}^\odot & \mathcal{O}_{\text{xy'}} & \mathcal{O}_{\text{yx'}} & \mathcal{O}_{\text{zy'}}\\
 46) & \text{P}_{\text{x}}^\text{c} & \text{P}_{\text{x}}^\odot & \text{P}_{\text{z}}^\text{s} & \text{P}_{\text{z}}^\odot & \text{P}_{\text{x'}}^\odot & \mathcal{O}_{\text{xy'}} & \mathcal{O}_{\text{yx'}} & \mathcal{O}_{\text{zy'}}\\
 47) & \text{P}_{\text{x}} & \text{P}_{\text{x}}^\text{c} & \text{P}_{\text{x}}^\text{s} & \text{P}_{\text{x}}^\odot & \text{P}_{\text{z}}^\text{c} & \text{P}_{\text{z}}^\text{s} & \text{P}_{\text{z'}}^\odot & \mathcal{O}_{\text{yz'}}\\
 48) & \text{P}_{\text{x}} & \text{P}_{\text{x}}^\text{c} & \text{P}_{\text{x}}^\text{s} & \text{P}_{\text{z}}^\text{c} & \text{P}_{\text{z}}^\text{s} & \text{P}_{\text{z'}}^\odot & \mathcal{O}_{\text{xy'}} & \mathcal{O}_{\text{yz'}}\\
 49) & \text{P}_{\text{x}}^\text{c} & \text{P}_{\text{x}}^\text{s} & \text{P}_{\text{x}}^\odot & \text{P}_{\text{z}}^\text{c} & \text{P}_{\text{z}}^\text{s} & \text{P}_{\text{z'}}^\odot & \mathcal{O}_{\text{xy'}} & \mathcal{O}_{\text{yz'}}\\
 50) & \text{P}_{\text{x}} & \text{P}_{\text{z}} & \text{P}_{\text{x}}^\odot & \text{P}_{\text{z}}^\odot & \text{P}_{\text{x'}}^\text{c} & \text{P}_{\text{z'}}^\text{s} & \mathcal{O}_{\text{xz'}} & \mathcal{O}_{\text{zx'}}\\
 51) & \text{P}_{\text{x}} & \text{P}_{\text{z}} & \text{P}_{\text{x'}}^\text{c} & \text{P}_{\text{z'}}^\text{s} & \mathcal{O}_{\text{xy'}} & \mathcal{O}_{\text{xz'}} & \mathcal{O}_{\text{zx'}} & \mathcal{O}_{\text{zy'}}\\
 52) & \text{P}_{\text{x}}^\odot & \text{P}_{\text{z}}^\odot & \text{P}_{\text{x'}}^\text{c} & \text{P}_{\text{z'}}^\text{s} & \mathcal{O}_{\text{xy'}} & \mathcal{O}_{\text{xz'}} & \mathcal{O}_{\text{zx'}} & \mathcal{O}_{\text{zy'}}\\
 53) & \text{I}^{^\text{c}} & \text{P}_{\text{x}} & \text{P}_{\text{x}}^\odot & \text{P}_{\text{y}}^\text{c} & \mathcal{O}_{\text{xx'}} & \mathcal{O}_{\text{xz'}} & \mathcal{O}_{\text{zx'}} & \mathcal{O}_{\text{zz'}}\\
 54) & \text{I}^{^\text{c}} & \text{P}_{\text{x}} & \text{P}_{\text{x}}^\odot & \text{P}_{\text{y'}}^\text{c} & \mathcal{O}_{\text{xx'}} & \mathcal{O}_{\text{xz'}} & \mathcal{O}_{\text{zx'}} & \mathcal{O}_{\text{zz'}}\\
 55) & \text{P}_{\text{x}} & \text{P}_{\text{x}}^\odot & \text{P}_{\text{y}}^\text{c} & \text{P}_{\text{y'}}^\text{c} & \mathcal{O}_{\text{xx'}} & \mathcal{O}_{\text{xz'}} & \mathcal{O}_{\text{zx'}} & \mathcal{O}_{\text{zz'}}\\
 56) & \text{I}^{^\text{c}} & \text{P}_{\text{x}} & \text{P}_{\text{y}}^\text{c} & \mathcal{O}_{\text{xx'}} & \mathcal{O}_{\text{xy'}} & \mathcal{O}_{\text{xz'}} & \mathcal{O}_{\text{zx'}} & \mathcal{O}_{\text{zz'}}\\
 57) & \text{I}^{^\text{c}} & \text{P}_{\text{x}} & \text{P}_{\text{y'}}^\text{c} & \mathcal{O}_{\text{xx'}} & \mathcal{O}_{\text{xy'}} & \mathcal{O}_{\text{xz'}} & \mathcal{O}_{\text{zx'}} & \mathcal{O}_{\text{zz'}}\\
 58) & \text{P}_{\text{x}} & \text{P}_{\text{y}}^\text{c} & \text{P}_{\text{y'}}^\text{c} & \mathcal{O}_{\text{xx'}} & \mathcal{O}_{\text{xy'}} & \mathcal{O}_{\text{xz'}} & \mathcal{O}_{\text{zx'}} & \mathcal{O}_{\text{zz'}}\\
 59) & \text{I}^{^\text{c}} & \text{P}_{\text{x}}^\odot & \text{P}_{\text{y}}^\text{c} & \mathcal{O}_{\text{xx'}} & \mathcal{O}_{\text{xy'}} & \mathcal{O}_{\text{xz'}} & \mathcal{O}_{\text{zx'}} & \mathcal{O}_{\text{zz'}}\\
 60) & \text{P}_{\text{x}}^\odot & \text{P}_{\text{y}}^\text{c} & \text{P}_{\text{y'}}^\text{c} & \mathcal{O}_{\text{xx'}} & \mathcal{O}_{\text{xy'}} & \mathcal{O}_{\text{xz'}} & \mathcal{O}_{\text{zx'}} & \mathcal{O}_{\text{zz'}}\\
 61) & \text{I}^{^\text{c}} & \text{P}_{\text{x}} & \text{P}_{\text{x}}^\odot & \text{P}_{\text{x'}}^\text{c} & \text{P}_{\text{x'}}^\text{s} & \text{P}_{\text{y'}}^\text{c} & \text{P}_{\text{z'}}^\text{c} & \text{P}_{\text{z'}}^\text{s} \\
 62) & \text{P}_{\text{x}} & \text{P}_{\text{x}}^\odot & \text{P}_{\text{y}}^\text{c} & \text{P}_{\text{x'}}^\text{c} & \text{P}_{\text{x'}}^\text{s} & \text{P}_{\text{y'}}^\text{c} & \text{P}_{\text{z'}}^\text{c} & \text{P}_{\text{z'}}^\text{s} \\
 63) & \text{I}^{^\text{c}} & \text{P}_{\text{x}} & \text{P}_{\text{y}}^\text{c} & \text{P}_{\text{x'}}^\text{c} & \text{P}_{\text{x'}}^\text{s} & \text{P}_{\text{z'}}^\text{c} & \text{P}_{\text{z'}}^\text{s} & \mathcal{O}_{\text{xy'}}\\
 64) & \text{I}^{^\text{c}} & \text{P}_{\text{x}}^\odot & \text{P}_{\text{y}}^\text{c} & \text{P}_{\text{x'}}^\text{c} & \text{P}_{\text{x'}}^\text{s} & \text{P}_{\text{z'}}^\text{c} & \text{P}_{\text{z'}}^\text{s} & \mathcal{O}_{\text{xy'}}\\
 65) & \text{I}^{^\text{c}} & \text{P}_{\text{x}}^\odot & \text{P}_{\text{x'}}^\text{c} & \text{P}_{\text{x'}}^\text{s} & \text{P}_{\text{y'}}^\text{c} & \text{P}_{\text{z'}}^\text{c} & \text{P}_{\text{z'}}^\text{s} & \mathcal{O}_{\text{xy'}}\\
 66) & \text{P}_{\text{x}}^\odot & \text{P}_{\text{y}}^\text{c} & \text{P}_{\text{x'}}^\text{c} & \text{P}_{\text{x'}}^\text{s} & \text{P}_{\text{y'}}^\text{c} & \text{P}_{\text{z'}}^\text{c} & \text{P}_{\text{z'}}^\text{s} & \mathcal{O}_{\text{xy'}}\\
 67) & \text{I}^{^\text{c}} & \text{I}^{\text{s}} & \text{P}_{\text{y}}^\text{c} & \text{P}_{\text{y}}^\text{s} & \text{P}_{\text{x'}}^\text{c} & \text{P}_{\text{z'}}^\text{s} & \mathcal{O}_{\text{xz'}} & \mathcal{O}_{\text{zx'}}\\
 68) & \text{I}^{^\text{c}} & \text{I}^{\text{s}} & \text{P}_{\text{x'}}^\text{c} & \text{P}_{\text{y'}}^\text{c} & \text{P}_{\text{y'}}^\text{s} & \text{P}_{\text{z'}}^\text{s} & \mathcal{O}_{\text{xz'}} & \mathcal{O}_{\text{zx'}}\\
 69) & \text{P}_{\text{y}}^\text{c} & \text{P}_{\text{y}}^\text{s} & \text{P}_{\text{x'}}^\text{c} & \text{P}_{\text{y'}}^\text{c} & \text{P}_{\text{y'}}^\text{s} & \text{P}_{\text{z'}}^\text{s} & \mathcal{O}_{\text{xz'}} & \mathcal{O}_{\text{zx'}}\\
 \\
\end{tabular}
\end{minipage}
\end{table*}
\end{ruledtabular}

\clearpage
\appendix
\section{Algorithm to check for completeness\label{algorithm}}
The following algorithm was designed by Lothar Tiator \cite{LotharPM} and was already applied in the paper \cite{PhysRevC.96.025210}.
It is used to check if a set of observables is able to resolve continuous as well as discrete ambiguities.
The starting point is a system of multivariate homogeneous polynomials $f_1(\vec{t}), \dots, f_n(\vec{t})$.
The input is a vector of $N$ complex amplitudes $t_i$.
Without loss of generality, the overall phase of the complex amplitudes is fixed by requiring $\Re(t_1)>0$ and $\Im(t_1)=0$.
In a next step an $N$ dimensional solution vector $\vec{s}$ is formed.
It consist of $2N-1$ randomly chosen prime numbers within a certain range.
These serve as values for the real and imaginary parts of the $t_i$.
Using prime numbers and increasing the range from which they are chosen should be capable of reducing the chance to land on a singularity in the solution space, where the “condition of equal magnitudes of relative phases"\cite{Nakayama} is met.

Finally, the polynomial system:
\begin{align}
    f_1(\vec{t}) &= g_1,\notag\\
    \vdots\\
    f_n(\vec{t}) &= g_n\notag,
\end{align}
is constructed, where $g_i = f_i(\vec{s})$ is a scalar quantity.
The function \textit{NSolve} from MATHEMATICA \cite{Mathematica} is employed to solve the algebraic system for the variables $t_1,\dots, t_N$.
According to the Wolfram Mathematica documentation \cite{mathematicaInternals}: “For systems of algebraic equations, NSolve computes a numerical Gröbner basis using an efficient monomial ordering, then uses eigensystem methods to extract numerical roots.”.

The system of polynomials is said to be complete, if only one solution is found which furthermore is equivalent to $\vec{s}$.

\begin{ruledtabular}
\begin{table}
\caption{Truly minimal complete sets, consisting of $2N=16$ observables, obtained for the example discussed in \cref{sec::algebraic_method}. The group I observables are not explicitly shown, as they are common to all complete sets. The used notation is analogous to Roberts/Oed \cite{RobertsOed}.\label{tab:YannicksResults}}
\begin{tabular}{CCCCCCCC}
\text{P}_{\text{x}} & \text{P}_{\text{z}} & \text{P}_{\text{x}}^{\odot} & \text{P}_{\text{z}}^{\odot} & \text{P}_{\text{z}}^{\text{s}} & \text{P}_{\text{x}}^{\text{c}} & \mathcal{O}_{\text{zz'}}^{\text{s}} & \mathcal{O}_{\text{xz'}}^{\text{c}} \\
\text{P}_{\text{x}} & \text{P}_{\text{z}} & \text{P}_{\text{x}}^{\odot} & \text{P}_{\text{z}}^{\odot} & \text{P}_{\text{z}}^{\text{s}} & \text{P}_{\text{x}}^{\text{c}} & \mathcal{O}_{\text{zx'}}^{\text{c}} & \mathcal{O}_{\text{xx'}}^{\text{s}} \\
\text{P}_{\text{x}} & \text{P}_{\text{z}} & \text{P}_{\text{x}}^{\odot} & \text{P}_{\text{z}}^{\odot} & \text{P}_{\text{z}}^{\text{s}} & \text{P}_{\text{x}}^{\text{c}} & \mathcal{O}_{\text{zz'}}^{\text{c}} & \mathcal{O}_{\text{xz'}}^{\text{s}} \\
\text{P}_{\text{x}} & \text{P}_{\text{z}} & \text{P}_{\text{x}}^{\odot} & \text{P}_{\text{z}}^{\odot} & \text{P}_{\text{z}}^{\text{s}} & \text{P}_{\text{x}}^{\text{c}} & \mathcal{O}_{\text{zx'}}^{\text{s}} & \mathcal{O}_{\text{xx'}}^{\text{c}} \\
\text{P}_{\text{x}} & \text{P}_{\text{z}} & \text{P}_{\text{x}}^{\odot} & \text{P}_{\text{z}}^{\odot} & \text{P}_{\text{z}}^{\text{c}} & \text{P}_{\text{x}}^{\text{s}} & \mathcal{O}_{\text{zz'}}^{\text{s}} & \mathcal{O}_{\text{xz'}}^{\text{c}} \\
\text{P}_{\text{x}} & \text{P}_{\text{z}} & \text{P}_{\text{x}}^{\odot} & \text{P}_{\text{z}}^{\odot} & \text{P}_{\text{z}}^{\text{c}} & \text{P}_{\text{x}}^{\text{s}} & \mathcal{O}_{\text{zx'}}^{\text{c}} & \mathcal{O}_{\text{xx'}}^{\text{s}} \\
\text{P}_{\text{x}} & \text{P}_{\text{z}} & \text{P}_{\text{x}}^{\odot} & \text{P}_{\text{z}}^{\odot} & \text{P}_{\text{z}}^{\text{c}} & \text{P}_{\text{x}}^{\text{s}} & \mathcal{O}_{\text{zz'}}^{\text{c}} & \mathcal{O}_{\text{xz'}}^{\text{s}} \\
\text{P}_{\text{x}} & \text{P}_{\text{z}} & \text{P}_{\text{x}}^{\odot} & \text{P}_{\text{z}}^{\odot} & \text{P}_{\text{z}}^{\text{c}} & \text{P}_{\text{x}}^{\text{s}} & \mathcal{O}_{\text{zx'}}^{\text{s}} & \mathcal{O}_{\text{xx'}}^{\text{c}} \\
\text{P}_{\text{x}} & \text{P}_{\text{z}} & \text{P}_{\text{x}}^{\odot} & \text{P}_{\text{z}}^{\odot} & \mathcal{O}_{\text{zy'}}^{\text{s}} & \mathcal{O}_{\text{xy'}}^{\text{c}} & \mathcal{O}_{\text{zz'}}^{\text{s}} & \mathcal{O}_{\text{xz'}}^{\text{c}} \\
\text{P}_{\text{x}} & \text{P}_{\text{z}} & \text{P}_{\text{x}}^{\odot} & \text{P}_{\text{z}}^{\odot} & \mathcal{O}_{\text{zy'}}^{\text{s}} & \mathcal{O}_{\text{xy'}}^{\text{c}} & \mathcal{O}_{\text{zx'}}^{\text{c}} & \mathcal{O}_{\text{xx'}}^{\text{s}} \\
\text{P}_{\text{x}} & \text{P}_{\text{z}} & \text{P}_{\text{x}}^{\odot} & \text{P}_{\text{z}}^{\odot} & \mathcal{O}_{\text{zy'}}^{\text{s}} & \mathcal{O}_{\text{xy'}}^{\text{c}} & \mathcal{O}_{\text{zz'}}^{\text{c}} & \mathcal{O}_{\text{xz'}}^{\text{s}} \\
\text{P}_{\text{x}} & \text{P}_{\text{z}} & \text{P}_{\text{x}}^{\odot} & \text{P}_{\text{z}}^{\odot} & \mathcal{O}_{\text{zy'}}^{\text{s}} & \mathcal{O}_{\text{xy'}}^{\text{c}} & \mathcal{O}_{\text{zx'}}^{\text{s}} & \mathcal{O}_{\text{xx'}}^{\text{c}} \\
\text{P}_{\text{x}} & \text{P}_{\text{z}} & \text{P}_{\text{x}}^{\odot} & \text{P}_{\text{z}}^{\odot} & \mathcal{O}_{\text{zy'}}^{\text{c}} & \mathcal{O}_{\text{xy'}}^{\text{s}} & \mathcal{O}_{\text{zz'}}^{\text{s}} & \mathcal{O}_{\text{xz'}}^{\text{c}} \\
\text{P}_{\text{x}} & \text{P}_{\text{z}} & \text{P}_{\text{x}}^{\odot} & \text{P}_{\text{z}}^{\odot} & \mathcal{O}_{\text{zy'}}^{\text{c}} & \mathcal{O}_{\text{xy'}}^{\text{s}} & \mathcal{O}_{\text{zx'}}^{\text{c}} & \mathcal{O}_{\text{xx'}}^{\text{s}} \\
\text{P}_{\text{x}} & \text{P}_{\text{z}} & \text{P}_{\text{x}}^{\odot} & \text{P}_{\text{z}}^{\odot} & \mathcal{O}_{\text{zy'}}^{\text{c}} & \mathcal{O}_{\text{xy'}}^{\text{s}} & \mathcal{O}_{\text{zz'}}^{\text{c}} & \mathcal{O}_{\text{xz'}}^{\text{s}} \\
\text{P}_{\text{x}} & \text{P}_{\text{z}} & \text{P}_{\text{x}}^{\odot} & \text{P}_{\text{z}}^{\odot} & \mathcal{O}_{\text{zy'}}^{\text{c}} & \mathcal{O}_{\text{xy'}}^{\text{s}} & \mathcal{O}_{\text{zx'}}^{\text{s}} & \mathcal{O}_{\text{xx'}}^{\text{c}} \\
\text{P}_{\text{z}}^{\text{s}} & \text{P}_{\text{x}}^{\text{c}} & \mathcal{O}_{\text{xy'}} & \mathcal{O}_{\text{zy'}} & \mathcal{O}_{\text{xy'}}^{\odot} & \mathcal{O}_{\text{zy'}}^{\odot} & \mathcal{O}_{\text{zz'}}^{\text{s}} & \mathcal{O}_{\text{xz'}}^{\text{c}} \\
\text{P}_{\text{z}}^{\text{s}} & \text{P}_{\text{x}}^{\text{c}} & \mathcal{O}_{\text{xy'}} & \mathcal{O}_{\text{zy'}} & \mathcal{O}_{\text{xy'}}^{\odot} & \mathcal{O}_{\text{zy'}}^{\odot} & \mathcal{O}_{\text{zx'}}^{\text{c}} & \mathcal{O}_{\text{xx'}}^{\text{s}} \\
\text{P}_{\text{z}}^{\text{s}} & \text{P}_{\text{x}}^{\text{c}} & \mathcal{O}_{\text{xy'}} & \mathcal{O}_{\text{zy'}} & \mathcal{O}_{\text{xy'}}^{\odot} & \mathcal{O}_{\text{zy'}}^{\odot} & \mathcal{O}_{\text{zz'}}^{\text{c}} & \mathcal{O}_{\text{xz'}}^{\text{s}} \\
\text{P}_{\text{z}}^{\text{s}} & \text{P}_{\text{x}}^{\text{c}} & \mathcal{O}_{\text{xy'}} & \mathcal{O}_{\text{zy'}} & \mathcal{O}_{\text{xy'}}^{\odot} & \mathcal{O}_{\text{zy'}}^{\odot} & \mathcal{O}_{\text{zx'}}^{\text{s}} & \mathcal{O}_{\text{xx'}}^{\text{c}} \\
\text{P}_{\text{z}}^{\text{c}} & \text{P}_{\text{x}}^{\text{s}} & \mathcal{O}_{\text{xy'}} & \mathcal{O}_{\text{zy'}} & \mathcal{O}_{\text{xy'}}^{\odot} & \mathcal{O}_{\text{zy'}}^{\odot} & \mathcal{O}_{\text{zz'}}^{\text{s}} & \mathcal{O}_{\text{xz'}}^{\text{c}} \\
\text{P}_{\text{z}}^{\text{c}} & \text{P}_{\text{x}}^{\text{s}} & \mathcal{O}_{\text{xy'}} & \mathcal{O}_{\text{zy'}} & \mathcal{O}_{\text{xy'}}^{\odot} & \mathcal{O}_{\text{zy'}}^{\odot} & \mathcal{O}_{\text{zx'}}^{\text{c}} & \mathcal{O}_{\text{xx'}}^{\text{s}} \\
\text{P}_{\text{z}}^{\text{c}} & \text{P}_{\text{x}}^{\text{s}} & \mathcal{O}_{\text{xy'}} & \mathcal{O}_{\text{zy'}} & \mathcal{O}_{\text{xy'}}^{\odot} & \mathcal{O}_{\text{zy'}}^{\odot} & \mathcal{O}_{\text{zz'}}^{\text{c}} & \mathcal{O}_{\text{xz'}}^{\text{s}} \\
\text{P}_{\text{z}}^{\text{c}} & \text{P}_{\text{x}}^{\text{s}} & \mathcal{O}_{\text{xy'}} & \mathcal{O}_{\text{zy'}} & \mathcal{O}_{\text{xy'}}^{\odot} & \mathcal{O}_{\text{zy'}}^{\odot} & \mathcal{O}_{\text{zx'}}^{\text{s}} & \mathcal{O}_{\text{xx'}}^{\text{c}} \\
\mathcal{O}_{\text{xy'}} & \mathcal{O}_{\text{zy'}} & \mathcal{O}_{\text{xy'}}^{\odot} & \mathcal{O}_{\text{zy'}}^{\odot} & \mathcal{O}_{\text{zy'}}^{\text{s}} & \mathcal{O}_{\text{xy'}}^{\text{c}} & \mathcal{O}_{\text{zz'}}^{\text{s}} & \mathcal{O}_{\text{xz'}}^{\text{c}} \\
\mathcal{O}_{\text{xy'}} & \mathcal{O}_{\text{zy'}} & \mathcal{O}_{\text{xy'}}^{\odot} & \mathcal{O}_{\text{zy'}}^{\odot} & \mathcal{O}_{\text{zy'}}^{\text{s}} & \mathcal{O}_{\text{xy'}}^{\text{c}} & \mathcal{O}_{\text{zx'}}^{\text{c}} & \mathcal{O}_{\text{xx'}}^{\text{s}} \\
\mathcal{O}_{\text{xy'}} & \mathcal{O}_{\text{zy'}} & \mathcal{O}_{\text{xy'}}^{\odot} & \mathcal{O}_{\text{zy'}}^{\odot} & \mathcal{O}_{\text{zy'}}^{\text{s}} & \mathcal{O}_{\text{xy'}}^{\text{c}} & \mathcal{O}_{\text{zz'}}^{\text{c}} & \mathcal{O}_{\text{xz'}}^{\text{s}} \\
\mathcal{O}_{\text{xy'}} & \mathcal{O}_{\text{zy'}} & \mathcal{O}_{\text{xy'}}^{\odot} & \mathcal{O}_{\text{zy'}}^{\odot} & \mathcal{O}_{\text{zy'}}^{\text{s}} & \mathcal{O}_{\text{xy'}}^{\text{c}} & \mathcal{O}_{\text{zx'}}^{\text{s}} & \mathcal{O}_{\text{xx'}}^{\text{c}} \\
\mathcal{O}_{\text{xy'}} & \mathcal{O}_{\text{zy'}} & \mathcal{O}_{\text{xy'}}^{\odot} & \mathcal{O}_{\text{zy'}}^{\odot} & \mathcal{O}_{\text{zy'}}^{\text{c}} & \mathcal{O}_{\text{xy'}}^{\text{s}} & \mathcal{O}_{\text{zz'}}^{\text{s}} & \mathcal{O}_{\text{xz'}}^{\text{c}} \\
\mathcal{O}_{\text{xy'}} & \mathcal{O}_{\text{zy'}} & \mathcal{O}_{\text{xy'}}^{\odot} & \mathcal{O}_{\text{zy'}}^{\odot} & \mathcal{O}_{\text{zy'}}^{\text{c}} & \mathcal{O}_{\text{xy'}}^{\text{s}} & \mathcal{O}_{\text{zx'}}^{\text{c}} & \mathcal{O}_{\text{xx'}}^{\text{s}} \\
\mathcal{O}_{\text{xy'}} & \mathcal{O}_{\text{zy'}} & \mathcal{O}_{\text{xy'}}^{\odot} & \mathcal{O}_{\text{zy'}}^{\odot} & \mathcal{O}_{\text{zy'}}^{\text{c}} & \mathcal{O}_{\text{xy'}}^{\text{s}} & \mathcal{O}_{\text{zz'}}^{\text{c}} & \mathcal{O}_{\text{xz'}}^{\text{s}} \\
\mathcal{O}_{\text{xy'}} & \mathcal{O}_{\text{zy'}} & \mathcal{O}_{\text{xy'}}^{\odot} & \mathcal{O}_{\text{zy'}}^{\odot} & \mathcal{O}_{\text{zy'}}^{\text{c}} & \mathcal{O}_{\text{xy'}}^{\text{s}} & \mathcal{O}_{\text{zx'}}^{\text{s}} & \mathcal{O}_{\text{xx'}}^{\text{c}}
\end{tabular}
\end{table}
\end{ruledtabular}

\begin{table*}[h]
\caption{Definition of the 64 $\Gamma$-matrices in terms of the well known Pauli matrices in combination with the Kronecker product. The gray shaded cells within the column "Shape-class" correspond to the non-zero matrix entries.}\label{definition_gamma_matrices}
\begin{tabular}{CCC}
    \hline\hline
    \Gamma\text{-matrices} & \multicolumn{1}{c}{Definition / 2} & \text{Shape-class}\\
    \midrule
\Gamma^\text{I}_1 & 2\cdot(\sigma^3 \otimes I_2 \otimes I_2) & \multirow{8}{*}{\includegraphics[width=3.5cm]{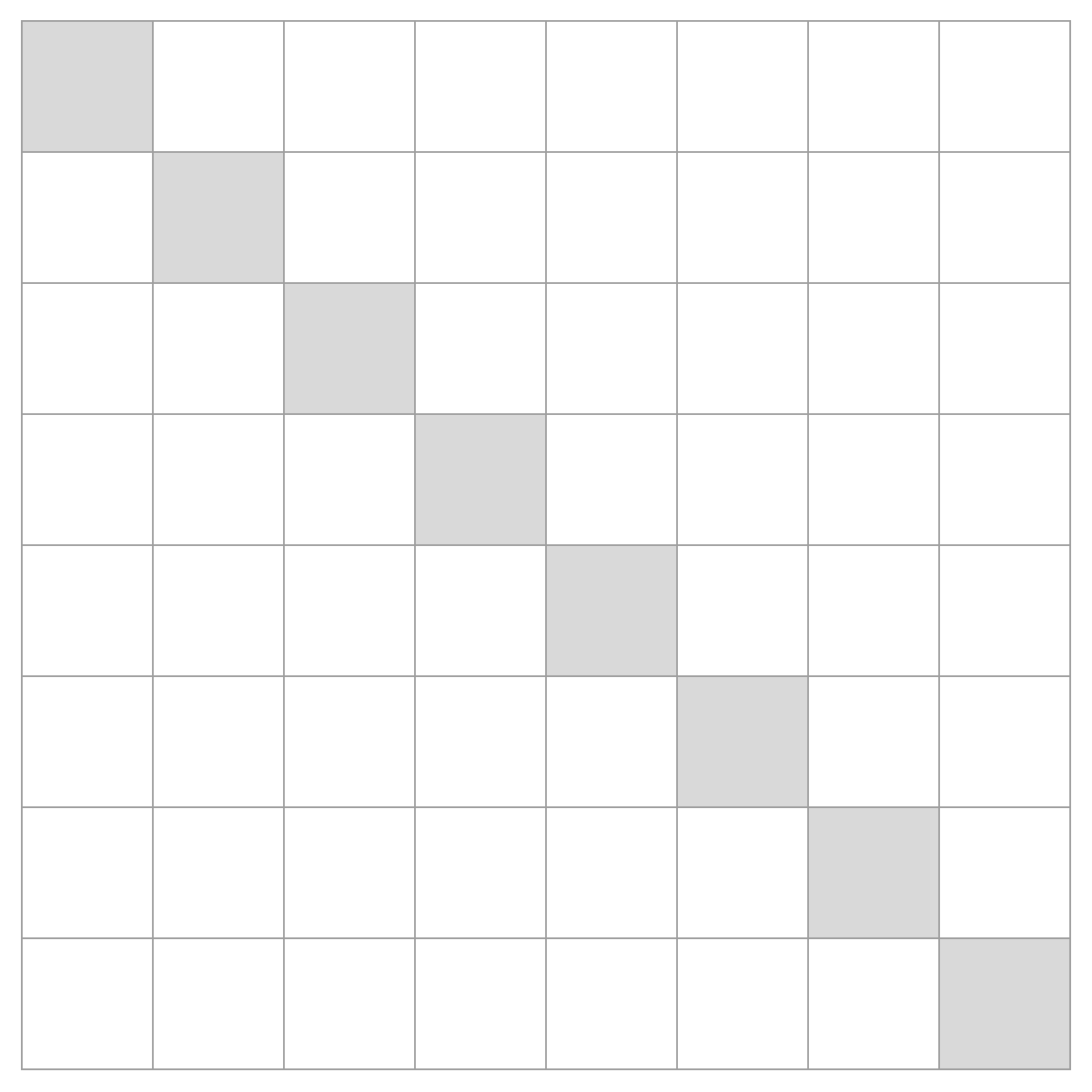}}\\
\Gamma^\text{I}_2 & 2\cdot(I_2 \otimes \sigma^3 \otimes I_2) & \\
\Gamma^\text{I}_3 & 2\cdot(I_2 \otimes I_2 \otimes \sigma^3) & \\
\Gamma^\text{I}_4 & 2\cdot(\sigma^3 \otimes \sigma^3 \otimes \sigma^3) & \\
\Gamma^\text{I}_5 & 2\cdot(I_2 \otimes \sigma^3 \otimes \sigma^3) & \\
\Gamma^\text{I}_6 & 2\cdot(\sigma^3 \otimes I_2 \otimes \sigma^3) & \\
\Gamma^\text{I}_7 & 2\cdot(\sigma^3 \otimes \sigma^3 \otimes I_2) & \\
\Gamma^\text{I}_8 & 2\cdot(I_2 \otimes I_2 \otimes I_2) & \\
    \midrule
\Gamma^\text{II}_{c1} & I_2 \otimes \sigma^1 \otimes I_2 & \multirow{8}{*}{\includegraphics[width=3.5cm]{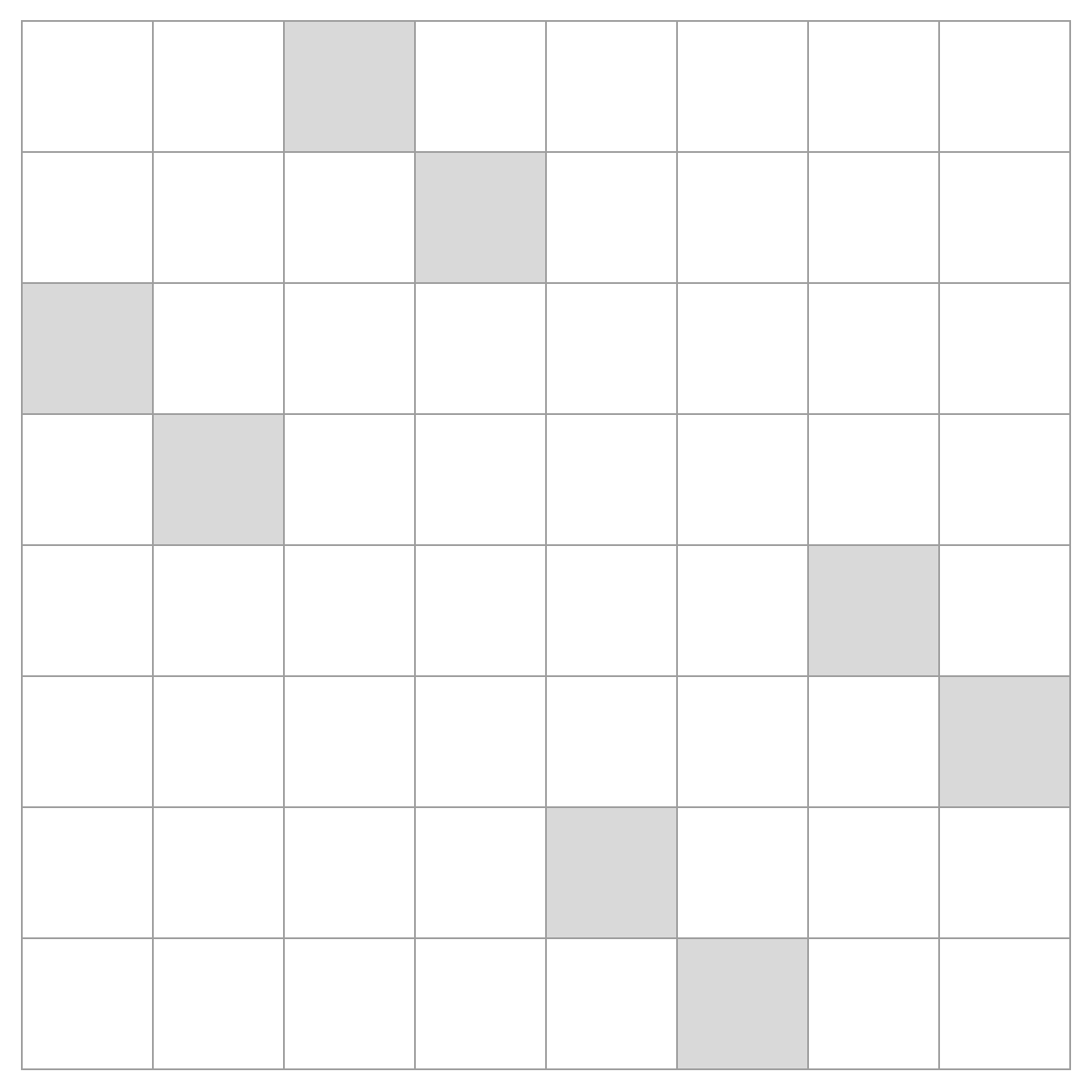}}\\
\Gamma^\text{II}_{c2} & \sigma^3 \otimes \sigma^1 \otimes I_2 & \\
\Gamma^\text{II}_{c3} & I_2 \otimes \sigma^1 \otimes \sigma^3 & \\
\Gamma^\text{II}_{c4} & \sigma^3 \otimes \sigma^1 \otimes \sigma^3 & \\
\Gamma^\text{II}_{s1} & -I_2 \otimes \sigma^2 \otimes I_2 & \\
\Gamma^\text{II}_{s2} & -\sigma^3 \otimes \sigma^2 \otimes I_2 & \\
\Gamma^\text{II}_{s3} & -I_2 \otimes \sigma^2 \otimes \sigma^3 & \\
\Gamma^\text{II}_{s4} & -\sigma^3 \otimes \sigma^2 \otimes \sigma^3 & \\
    \midrule
\Gamma^\text{III}_{c1} & I_2 \otimes I_2 \otimes \sigma^1 & \multirow{8}{*}{\includegraphics[width=3.5cm]{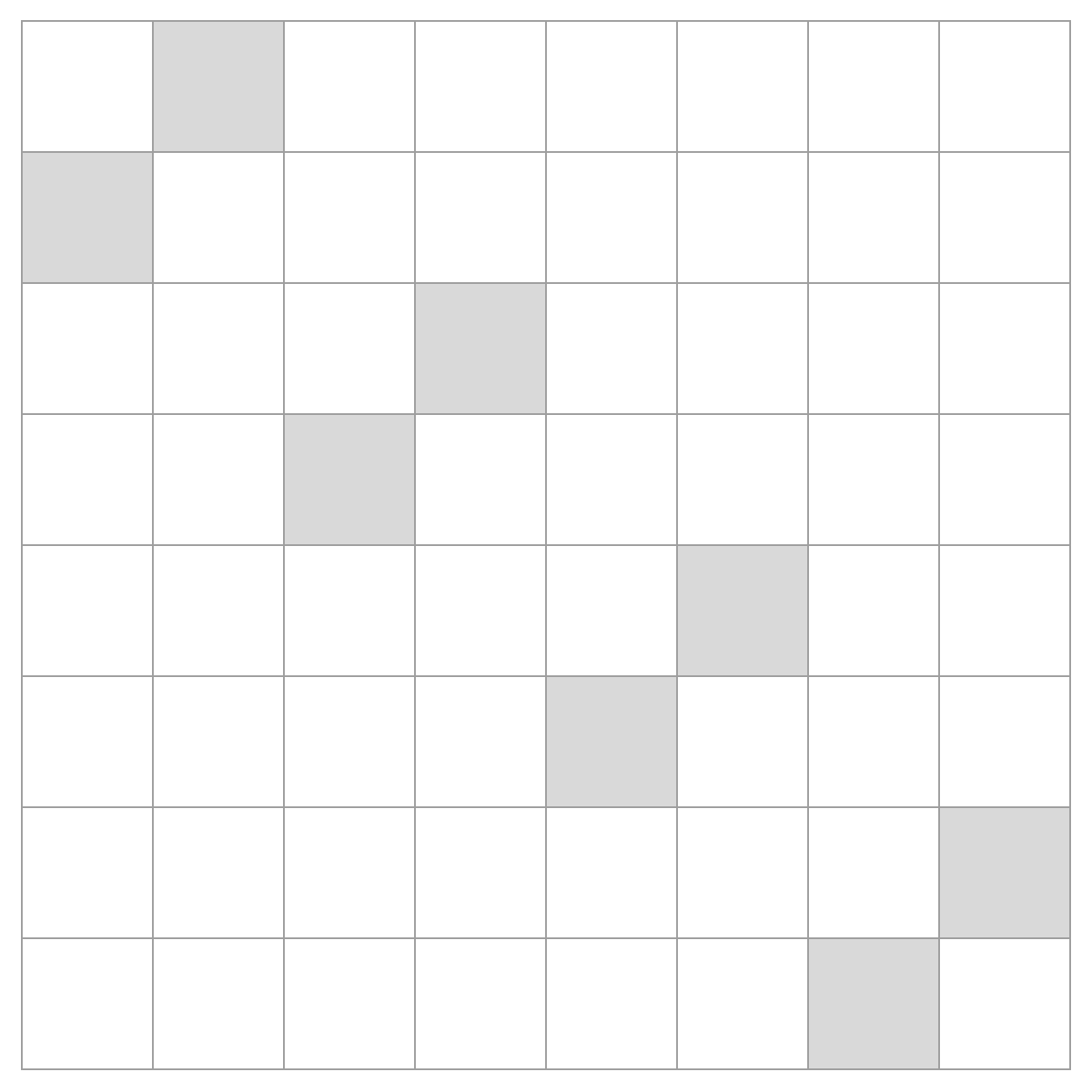}}\\
\Gamma^\text{III}_{c2} & \sigma^3 \otimes I_2 \otimes \sigma^1 & \\
\Gamma^\text{III}_{c3} & I_2 \otimes \sigma^3 \otimes \sigma^1 & \\
\Gamma^\text{III}_{c4} & \sigma^3 \otimes \sigma^3 \otimes \sigma^1 & \\
\Gamma^\text{III}_{s1} & -I_2 \otimes I_2 \otimes \sigma^2 & \\
\Gamma^\text{III}_{s2} & -\sigma^3 \otimes I_2 \otimes \sigma^2 & \\
\Gamma^\text{III}_{s3} & -I_2 \otimes \sigma^3 \otimes \sigma^2 & \\
\Gamma^\text{III}_{s4} & -\sigma^3 \otimes \sigma^3 \otimes \sigma^2 & \\
    \midrule
\Gamma^\text{IV}_{c1} & I_2 \otimes \sigma^1 \otimes \sigma^1 & \multirow{8}{*}{\includegraphics[width=3.5cm]{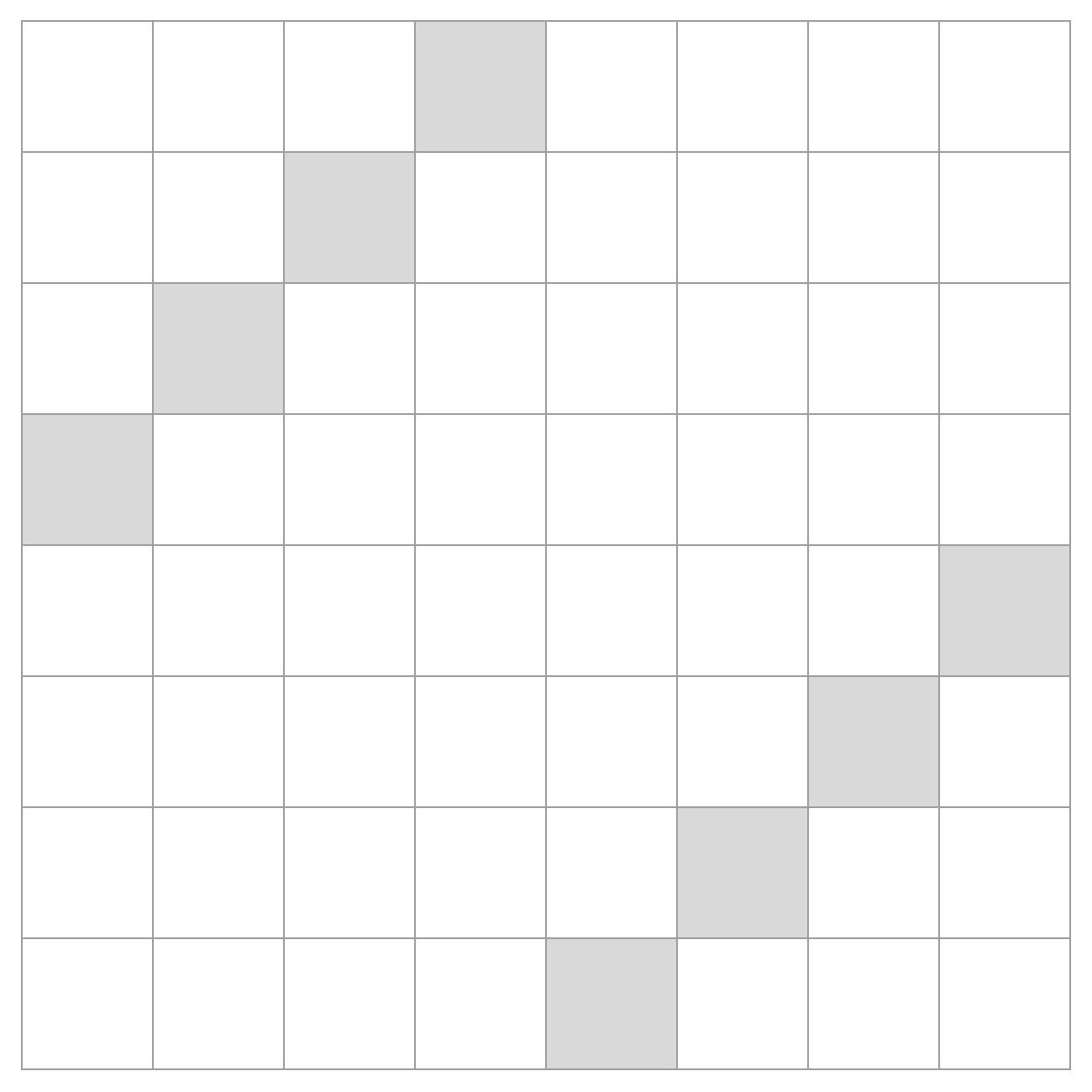}}\\
\Gamma^\text{IV}_{c2} & \sigma^3 \otimes \sigma^1 \otimes \sigma^1 & \\
\Gamma^\text{IV}_{c3} & -I_2 \otimes \sigma^2 \otimes \sigma^2 & \\
\Gamma^\text{IV}_{c4} & -\sigma^3 \otimes \sigma^2 \otimes \sigma^2 & \\
\Gamma^\text{IV}_{s1} & -I_2 \otimes \sigma^2 \otimes \sigma^1 & \\
\Gamma^\text{IV}_{s2} & -\sigma^3 \otimes \sigma^2 \otimes \sigma^1 & \\
\Gamma^\text{IV}_{s3} & -I_2 \otimes \sigma^1 \otimes \sigma^2 & \\
\Gamma^\text{IV}_{s4} & -\sigma^3 \otimes \sigma^1 \otimes \sigma^2 & \\
    \hline\hline
\end{tabular}
\qquad\qquad\qquad
\begin{tabular}{CCC}
    \hline\hline
    \Gamma\text{-matrices} & \multicolumn{1}{c}{Definition / 2} & \text{Shape-class}\\
    \midrule
\Gamma^\text{V}_{c1} & \sigma^1 \otimes I_2 \otimes I_2 & \multirow{8}{*}{\includegraphics[width=3.5cm]{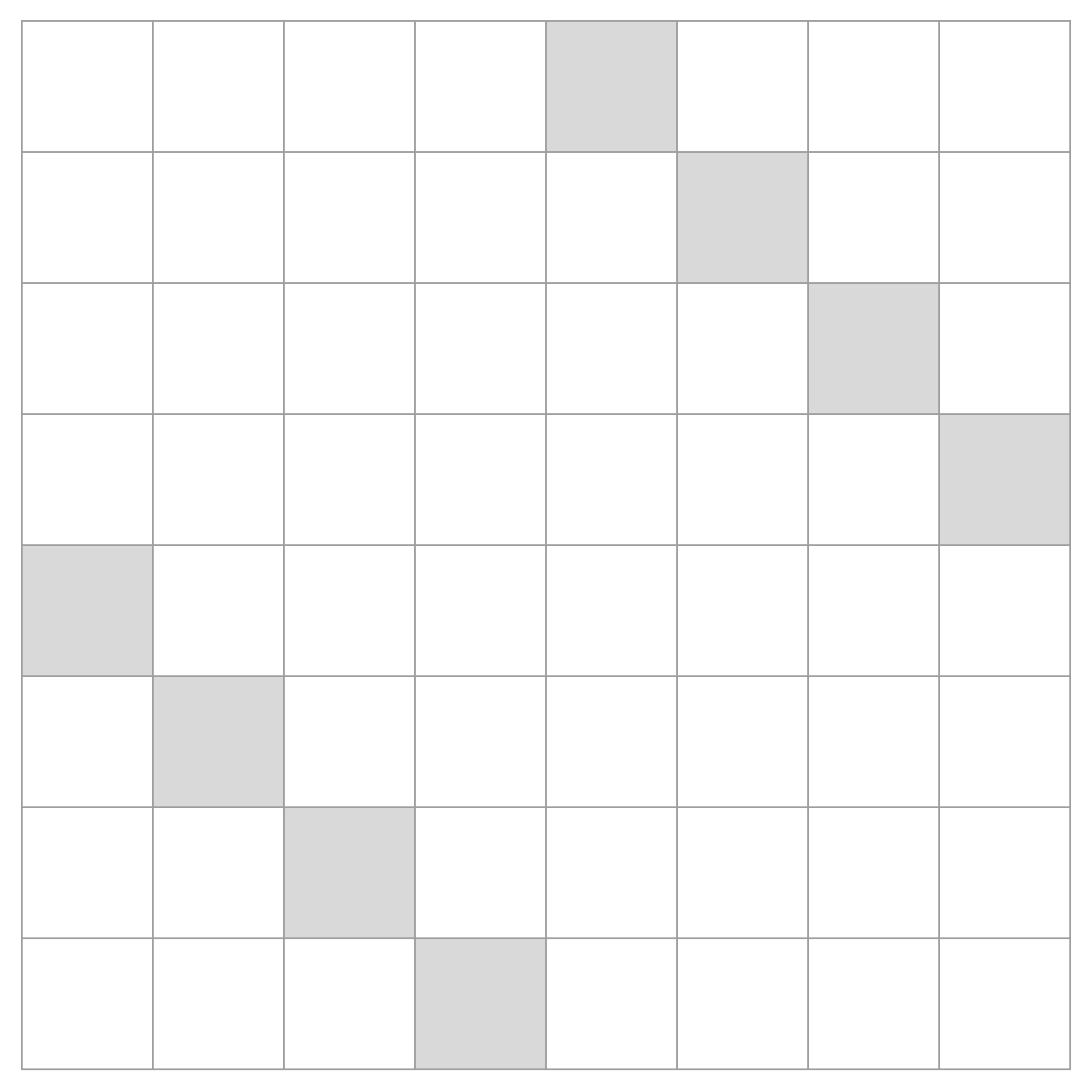}}\\
\Gamma^\text{V}_{c2} & \sigma^1 \otimes \sigma^3 \otimes I_2 & \\
\Gamma^\text{V}_{c3} & \sigma^1 \otimes I_2 \otimes \sigma^3 & \\
\Gamma^\text{V}_{c4} & \sigma^1 \otimes \sigma^3 \otimes \sigma^3 & \\
\Gamma^\text{V}_{s1} & -\sigma^2 \otimes I_2 \otimes I_2 & \\
\Gamma^\text{V}_{s2} & -\sigma^2 \otimes \sigma^3 \otimes I_2 & \\
\Gamma^\text{V}_{s3} & -\sigma^2 \otimes I_2 \otimes \sigma^3 & \\
\Gamma^\text{V}_{s4} & -\sigma^2 \otimes \sigma^3 \otimes \sigma^3 & \\
    \midrule
\Gamma^\text{VI}_{c1} & \sigma^1 \otimes \sigma^1 \otimes I_2 & \multirow{8}{*}{\includegraphics[width=3.5cm]{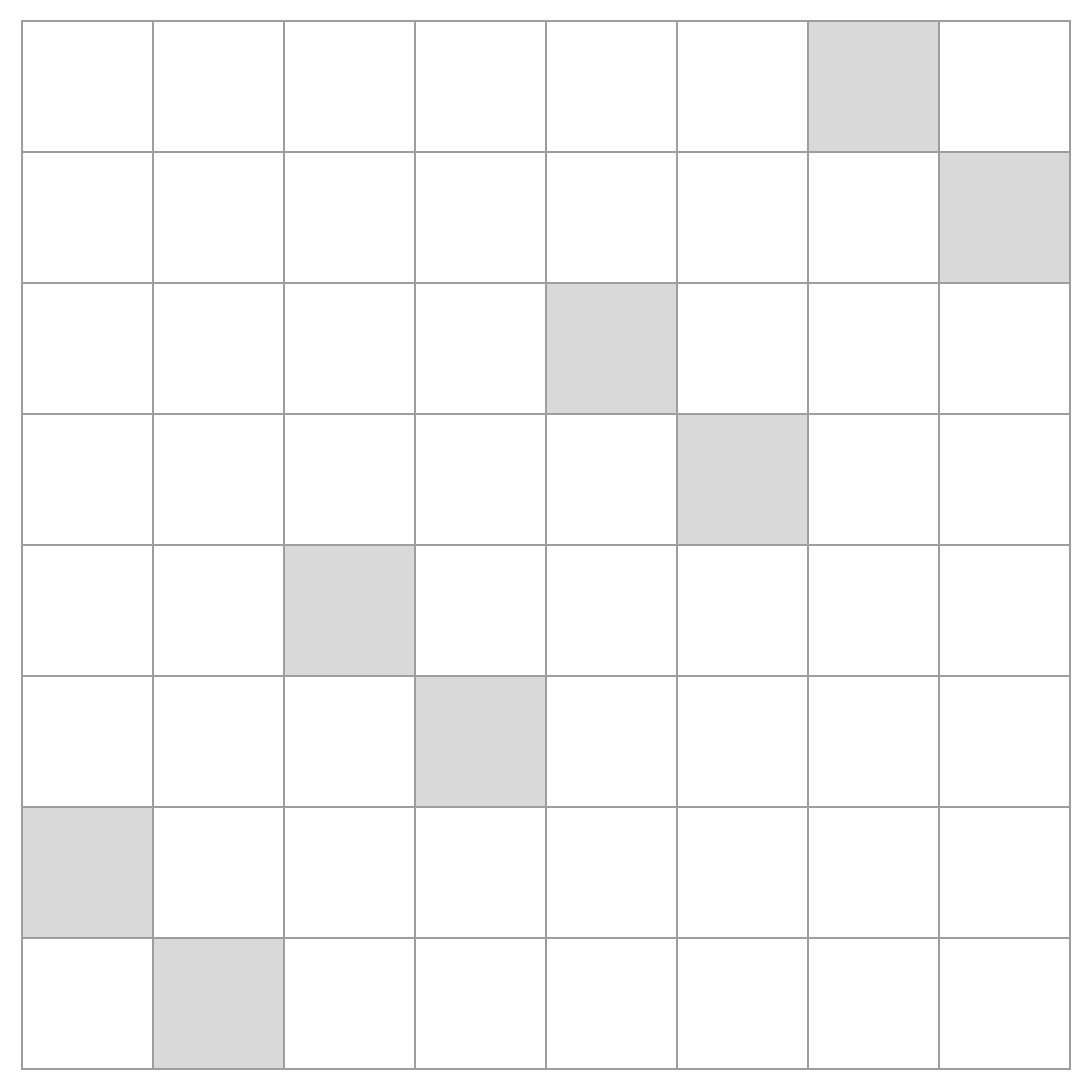}}\\
\Gamma^\text{VI}_{c2} & -\sigma^2 \otimes \sigma^2 \otimes I_2 & \\
\Gamma^\text{VI}_{c3} & \sigma^1 \otimes \sigma^1 \otimes \sigma^3 & \\
\Gamma^\text{VI}_{c4} & -\sigma^2 \otimes \sigma^2 \otimes \sigma^3 & \\
\Gamma^\text{VI}_{s1} & -\sigma^2 \otimes \sigma^1 \otimes I_2 & \\
\Gamma^\text{VI}_{s2} & -\sigma^1 \otimes \sigma^2 \otimes I_2 & \\
\Gamma^\text{VI}_{s3} & -\sigma^2 \otimes \sigma^1 \otimes \sigma^3 & \\
\Gamma^\text{VI}_{s4} & -\sigma^1 \otimes \sigma^2 \otimes \sigma^3 & \\
    \midrule
\Gamma^\text{VII}_{c1} & \sigma^1 \otimes I_2 \otimes \sigma^1 & \multirow{8}{*}{\includegraphics[width=3.5cm]{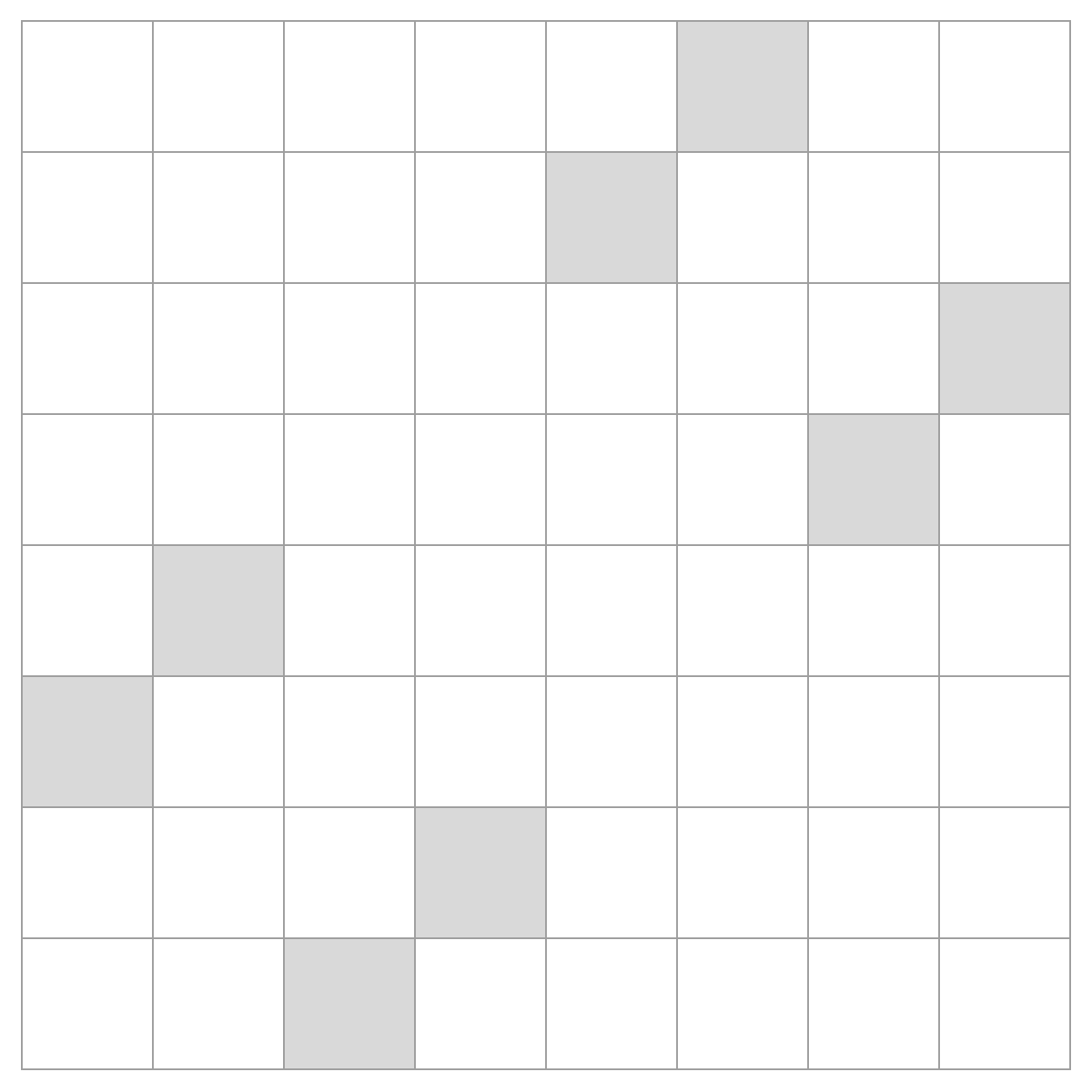}}\\
\Gamma^\text{VII}_{c2} & \sigma^1 \otimes \sigma^3 \otimes \sigma^1 & \\
\Gamma^\text{VII}_{c3} & -\sigma^2 \otimes I_2 \otimes \sigma^2 & \\
\Gamma^\text{VII}_{c4} & -\sigma^2 \otimes \sigma^3 \otimes \sigma^2 & \\
\Gamma^\text{VII}_{s1} & -\sigma^2 \otimes I_2 \otimes \sigma^1 & \\
\Gamma^\text{VII}_{s2} & -\sigma^2 \otimes \sigma^3 \otimes \sigma^1 & \\
\Gamma^\text{VII}_{s3} & -\sigma^1 \otimes I_2 \otimes \sigma^2 & \\
\Gamma^\text{VII}_{s4} & -\sigma^1 \otimes \sigma^3 \otimes \sigma^2 & \\
    \midrule
\Gamma^\text{VIII}_{c1} & \sigma^1 \otimes \sigma^1 \otimes \sigma^1 & \multirow{8}{*}{\includegraphics[width=3.5cm]{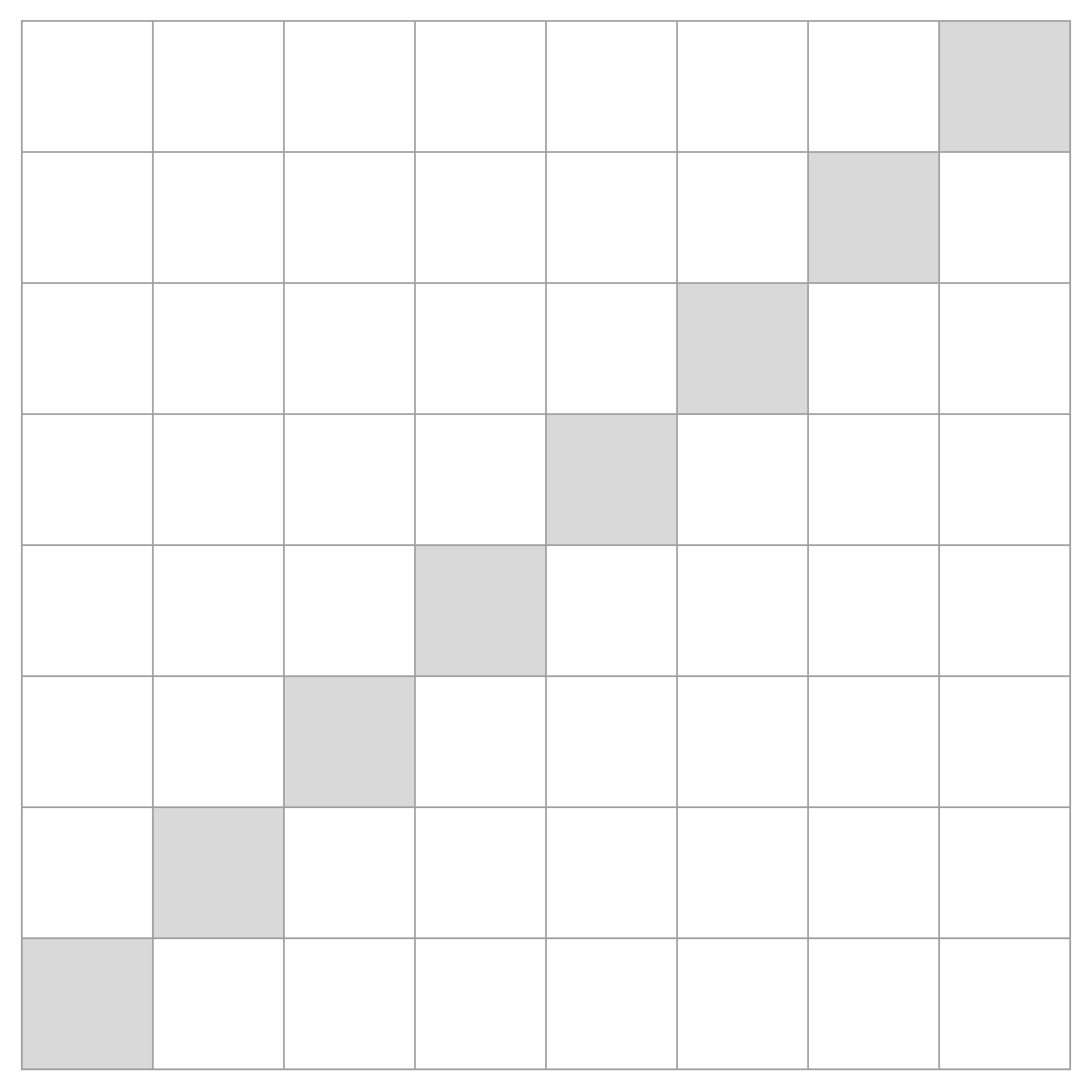}}\\
\Gamma^\text{VIII}_{c2} & -\sigma^2 \otimes \sigma^2 \otimes \sigma^1 & \\
\Gamma^\text{VIII}_{c3} & -\sigma^2 \otimes \sigma^1 \otimes \sigma^2 & \\
\Gamma^\text{VIII}_{c4} & -\sigma^1 \otimes \sigma^2 \otimes \sigma^2 & \\
\Gamma^\text{VIII}_{s1} & -\sigma^2 \otimes \sigma^1 \otimes \sigma^1 & \\
\Gamma^\text{VIII}_{s2} & -\sigma^1 \otimes \sigma^2 \otimes \sigma^1 & \\
\Gamma^\text{VIII}_{s3} & -\sigma^1 \otimes \sigma^1 \otimes \sigma^2 & \\
\Gamma^\text{VIII}_{s4} & \sigma^2 \otimes \sigma^2 \otimes \sigma^2 & \\
    \hline\hline
\end{tabular}
\end{table*}

\clearpage

\bibliography{paper}

\end{document}